\documentclass{article}
\usepackage {appendix}
\usepackage{amsthm}
\usepackage{PRIMEarxiv}
\usepackage{amsmath}
\usepackage[utf8]{inputenc} 
\usepackage[T1]{fontenc}    
\usepackage{hyperref}       
\usepackage{url}            
\usepackage{booktabs}       
\usepackage{amsfonts}       
\usepackage{nicefrac}       
\usepackage{microtype}      
\usepackage{lipsum}
\usepackage{fancyhdr}       
\usepackage{graphicx}       

\newtheorem{thm}{Theorem}

\graphicspath{{media/}}     

\title{A Logarithm Depth Quantum Converter: From One-hot Encoding to Binary Encoding.
\thanks{\textit{\underline{Citation}}: 
\textbf{Authors. Title. Pages.... DOI:000000/11111.}} 
}

\author{
  Bingren Chen, Hanqing Wu, Haomu Yuan, Lei Wu, Xin Li \\
  CCB fintech \\
  \texttt{bingren@mail.ustc.edu.cn} \\
}

\begin{document}
\maketitle

\begin{abstract}
  Within the quantum computing, there are two ways to encode a normalized vector $\{ \alpha_i \}$. They are one-hot encoding and binary coding. The one-hot encoding state is denoted as $\left | \psi_O^{(N)} \right \rangle=\sum_{i=0}^{N-1} \alpha_i \left |0 \right \rangle^{\otimes N-i-1} \left |1 \right \rangle \left |0 \right \rangle ^{\otimes i}$ and the binary encoding state is denoted as $\left | \psi_B^{(N)} \right \rangle=\sum_{i=0}^{N-1} \alpha_i \left |b_i \right \rangle$, where $b_i$ is interpreted in binary of $i$ as the tensor product sequence of qubit states. In this paper, we present a method converting between the one-hot encoding state and the binary encoding state by taking the Edick state as the transition state, where the Edick state is defined as $\left | \psi_E^{(N)} \right \rangle=\sum_{i=0}^{N-1} \alpha_i \left |0 \right \rangle^{\otimes N-i-1} \left |1 \right \rangle ^{\otimes i}$. Compared with the early work, our circuit achieves the exponential speedup with $O(\log^2 N)$ depth and $O(N)$ size.
  \end{abstract}

\section{Introduction}

Encoding normalized vectors into the amplitude of the initial quantum state is required in most quantum algorithms. Normally, there are two ways to encode the normalized vectors, which are one-hot encoding and binary encoding. For the vector $\{ \alpha_i \}$, the one-hot encoding state is denoted as $\left | \psi_O^{(N)} \right \rangle=\sum_{i=0}^{N-1} \alpha_i \left |0 \right \rangle^{\otimes N-i-1} \left |1 \right \rangle \left |0 \right \rangle ^{\otimes i}$ and the binary encoding state is denoted as $\left | \psi_B^{(N)} \right \rangle=\sum_{i=0}^{N-1} \alpha_i \left |b_i \right \rangle$, where $b_i$ is interpreted in binary of $i$ as the tensor product sequence of qubit states. The notation $N$ presents the length of the vector $\{ \alpha_i \}$. One-hot encoding state is often used to solve combinatorial optimization problems by using quantum approximate optimization algorithm \cite{farhi_quantum_2014, hadfield_quantum_2019} because the Ansatz is easier to design and uses fewer multi-qubit gates. While the binary encoding is more popular in the quantum Fourier algorithms such as HHL \cite{harrow_quantum_2009} and quantum linear discriminant analysis (QLDA) \cite{cong_quantum_2016}.

  When encoding an $N$-length vector, one-hot encoding needs $N$ qubits while  binary encoding needs $\log N$ qubits. Also, the uniform superposition states whose coefficients are $\alpha_i=1/\sqrt{N}$ for each $i$ are easier to prepare in binary coding by using the Hadamard gates whose depth is $O(1)$. By contrast, the preparation of W states (the uniform superposition states in one-hot encoding) requires an $O(\log N)$-depth scheme \cite{cruz_efficient_2019}.

  However, if we want to evolve the quantum state into a specific constrained state, the one-hot encoding may be easier to be implemented than the binary encoding. For example, some QAOA algorithms \cite{hadfield_quantum_2019, rebentrost_quantum_2018} apply one-hot encoding because it supports a simpler compilation of the mixer operator. Moreover, the one-hot encoding has higher fault tolerance than the binary encoding.

  Given  the vector $\{ \alpha_i \}$, the circuit to prepare $\left | \psi_O^{(N)} \right \rangle$ and $\left | \psi_B^{(N)} \right \rangle$ from $\left | 0 \right \rangle$ has $O(N)$ depth \cite{grover_creating_2002, mottonen_transformation_2005, iten_quantum_2016}. However, if the vector is unknown, how to design a converter scheme between $\left | \psi_O^{(N)} \right \rangle$ and $\left | \psi_B^{(N)} \right \rangle$ is a challenging problem. The converter plays an important role in a quantum algorithm using one-hot encoding at first and binary encoding at last. Some schemes \cite{plesch_efficient_2010, bartschi_deterministic_2019}use the converter as a intermediate step to implement the quantum compression on the symmetric pure state.

  In this paper, we present a new method converting between the one-hot encoding state and the binary encoding state by taking the Edick state as the transition state. The Edick state is defined as $\left | \psi_E^{(N)} \right \rangle=\sum_{i=0}^{N-1} \alpha_i \left |0 \right \rangle^{\otimes N-1-i} \left |1 \right \rangle ^{\otimes i}$.  The converter circuit depth is $O(\log^2 N)$ and the size is $O(N)$. Compared with the early work \cite{bartschi_deterministic_2019} whose depth and size is $O(N \log N)$, our scheme achieves an exponential speedup. Besides, with the help of some schemes \cite{abdullah-al-shafi_new_2017, das_reversible_2015, karkaj_binary_2017}, our solution can also realize the conversion between one-hot code and Grey code \cite{10.1145/360336.360343} in quantum encoding.

  The Edick state is inspired by the Dicke state \cite{bartschi_deterministic_2019}, which is denoted as $\left | D_k^{n} \right \rangle=\binom{n}{k}^{-\frac{1}{2}} \sum_{x \in \{0,1\}^n, wt(x)=k} \left |x \right \rangle$. We have, e.g. $\left | D_4^{2} \right \rangle=\frac{1}{\sqrt{6}} (\left |1100 \right \rangle+\left |1010 \right \rangle+\left |1001 \right \rangle+\left |0110 \right \rangle+\left |0101 \right \rangle+\left |0011 \right \rangle)$. Based on the idea from \cite{bartschi_deterministic_2019}, we use our converter to prepare one-hot encoding state and binary coding state for binomial distribution, which is denoted as $\left | \psi \right \rangle=\sum_{k=0}^{N-1} \sqrt{f(k, p)}\left | k \right \rangle$, where $f(k, p) = \binom {n}{k} p^k (1-p)^{n-k}$ is the probability density function of the binomial distribution with $\mathcal{B}(n, p)$. According to the central limit theorem, the binomial distribution $\mathcal{B}(n, p)$ tends to the normal distribution $\mathcal{N}(np, np(1-p))$ as $n$ increases. Therefore, this method can approximately prepare a normal distribution state, which is used in the some quantum finance applications \cite{rebentrost_quantum_2018, woerner_quantum_2019}.

  The paper is listed as bellow: In Sec. \ref{sec2}, we propose the exponential speedup scheme converting between the one-hot encoding state and the binary encoding. Sec. \ref{sec3} gives an application of the converter to prepare the binomial distribution state.
  Sec. \ref{sec5} compares our scheme with other related works.
  Sec. \ref{sec4} shows the numerical simulation of the converter and Sec. \ref{sec6} is the conclusion.

  \section{Conversion between the one-hot encoding state and the binary encoding state}
\label{sec2}

In this section, we present two schemes to prepare the one-hot encoding state $\left | \psi_O^{(N)} \right \rangle$ and the binary encoding state $\left | \psi_B^{(N)} \right \rangle$ from the Edick state $\left | \psi_E^{(N)} \right \rangle$. $\left | \psi_O^{(N)} \right \rangle$, $\left | \psi_B^{(N)} \right \rangle$ and $\left | \psi_E^{(N)} \right \rangle$ respectively include $N$, $\lceil \log N \rceil$ and $N-1$ qubits. It is easy to see the following theorem.

\begin{thm}
  If there exists two operators , $U_O$ and $U_B$, satisfying that $U_O (\left | \psi_E^{(N)} \right \rangle \otimes \left |1\right \rangle)= \left | \psi_O^{(N)} \right \rangle$ \footnote{$\left | \psi_E^{(N)} \right \rangle$ has $N-1$ qubits and $\left | \psi_O^{(N)} \right \rangle$ has $N$ qubits, so here we should tensor a qubit} and $U_B \left | \psi_E^{(N)} \right \rangle= \left | 0 \right \rangle ^{\otimes N-1-\lceil\log N\rceil}\left | \psi_B^{(N)} \right \rangle$ \footnote{$\left | \psi_E^{(N)} \right \rangle$ has $N-1$ qubits and $\left | \psi_B^{(N)} \right \rangle$ has $\lceil\log N\rceil$ qubits, so here we should tensor $N-1-\lceil\log N\rceil$ qubits}, then it is clear that $(U_B \otimes I) U_O^{-1} \left | \psi_O^{(N)} \right \rangle=\left | 0 \right \rangle ^{N-1-\lceil\log N\rceil}\left | \psi_B^{(N)} \right \rangle \left |1 \right \rangle$ and  $U_O (U_B^{-1} \otimes I) (\left | 0 \right \rangle ^{\otimes N-1-\lceil\log N\rceil}\left | \psi_B^{(N)} \right \rangle \left | 1 \right \rangle) =\left | \psi_O^{(N)} \right \rangle$ can transfer between the one-hot encoding state and the binary encoding.
\end{thm}

\subsection{$U_O^{(N)}$: From $\left | \psi_E^{(N)} \right \rangle$ to $\left | \psi_O^{(N)} \right \rangle$}

We use the idea of recursion to propose a $O(\log N)$ complexity scheme. The circuit is shown as Fig. \ref{f9} and Fig. \ref{f8}. A detailed example of $U_O^{(N)}$ is shown in Appendix \ref{a1}.

\begin{figure}[hbtp]
  \centering
  \includegraphics[width=0.6\textwidth,trim=50 0 0 0, clip]{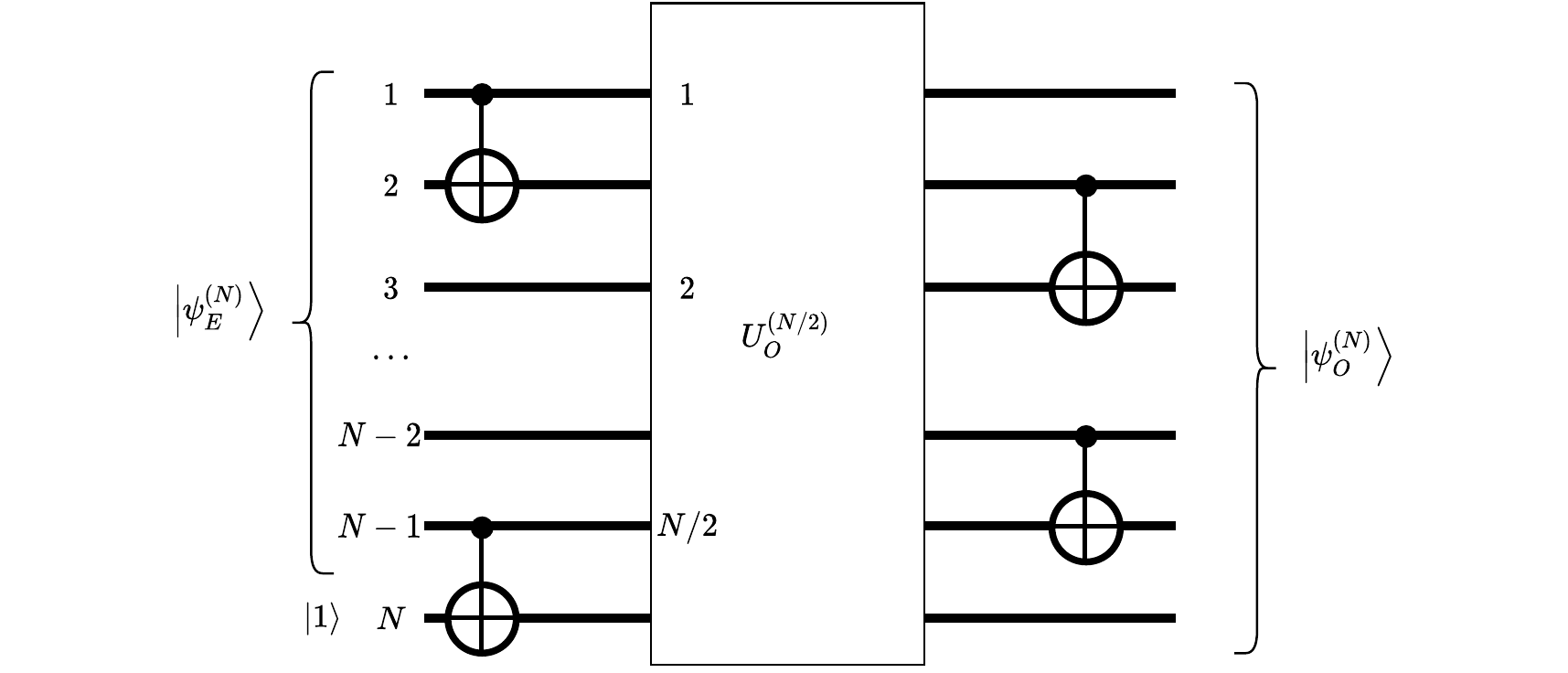}
  \caption{Edick to one-hot encoding state circuit $U_O^{(N)}$ when $N$ is even.}
  \label{f9}
\end{figure}

When $N$ is even, $\frac{N}{2}$ CNOT gates are applied with the odd qubits  \footnote{''Odd'' is for the qubit number but not for the index $i$.} as the control qubits and their next qubits as the target qubits. Then, the quantum state is
\begin{equation}
\begin{aligned}
 & CNOT^{\otimes \frac{N}{2}} (\left | \psi_E^{(N)}\right \rangle\otimes \left |1\right\rangle )= \\
 &\sum_{0 \leq i \leq N, i\%2=1} \alpha_i \left|0\right \rangle ^{\otimes N-i-1} \left |10\right \rangle ^{\otimes (i+1)/2}\\
 &+\sum_{0 \leq i \leq N, i\%2=0}\alpha_i \left|0\right \rangle ^{\otimes N-i-1} \left|1\right\rangle\left |10\right \rangle ^{\otimes i/2}.
\end{aligned}
\end{equation}

Given that we have prepared a smaller circuit $U_O^{(N/2)}$ and have applied it on all odd qubits, then according to the definition of $U_O$ the state becomes
\begin{equation}
\begin{aligned}
  \sum_{0 \leq i \leq N, i\%2=1} \alpha_i \left|0\right \rangle ^{\otimes N-i-1} \left |1\right \rangle \left |0\right \rangle ^{\otimes i} \\+\sum_{0 \leq i \leq N, i\%2=0}\alpha_i \left|0\right \rangle ^{\otimes N-i-1} \left|11\right\rangle\left |0\right \rangle ^{\otimes (i-1)}.
\end{aligned}
\end{equation}

Finally, $\frac{N}{2}-1$ CNOT gates are applied with all even qubits as the control qubits and their next qubits as the target qubits, and the quantum state is transformed to $\left |\psi_O^{N} \right \rangle$. It should be noted that $U_O^{(2)}$ is equal to the CNOT gate:
\begin{equation}
  (\alpha_0 \left | 0 \right \rangle+ \alpha_1 \left | 1 \right \rangle)\left |1 \right \rangle \rightarrow \alpha_0 \left | 01 \right \rangle+ \alpha_1 \left | 10 \right \rangle.
\end{equation}

\begin{figure}[hbtp]
  \centering
  \includegraphics[width=0.6\textwidth,trim=50 0 0 0, clip]{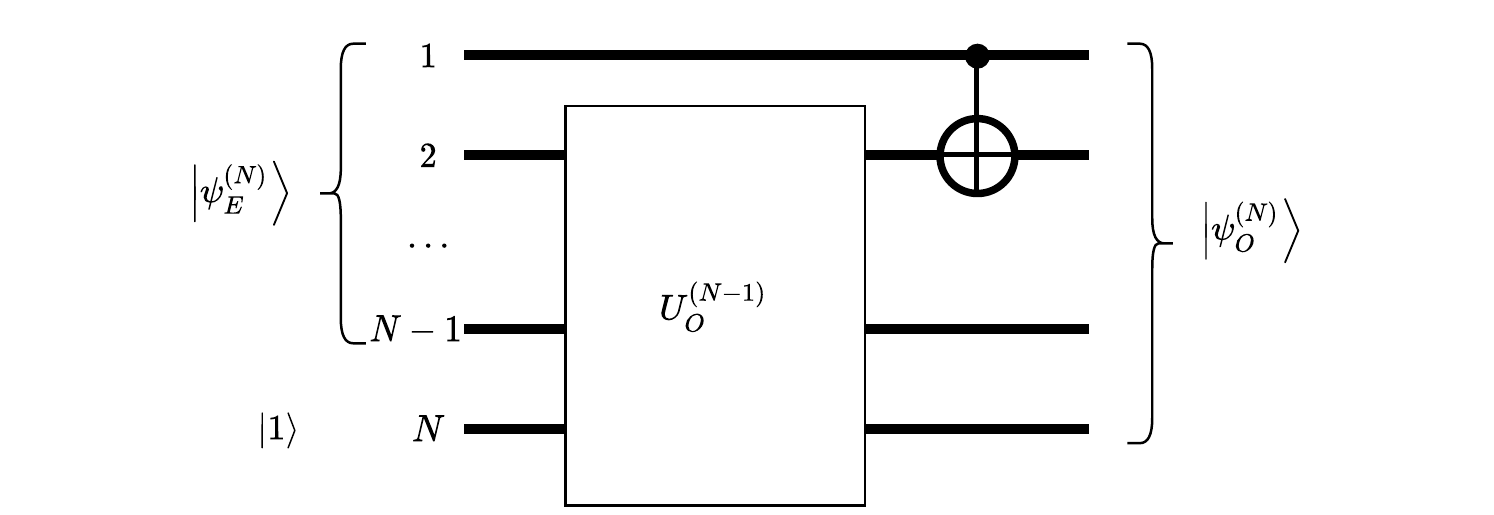}
  \caption{Edick to one-hot encoding state circuit $U_O^{(N)}$ when $N$ is odd.}
  \label{f8}
\end{figure}

When $N$ is odd, the initial state can be rewritten as
\begin{equation}
  \alpha_{N-1} \left |1\right \rangle \left |1\right \rangle^{\otimes N-1}+\left |0\right \rangle \left |\psi_E^{N-1} \right \rangle\left |1\right \rangle.
\end{equation}

After applying $U_O^{(N-1)}$, the state is
\begin{equation}
  \alpha_{N-1}\left |11\right \rangle \left |0\right \rangle^{\otimes N-2}+\left |0\right \rangle \left |\psi_O^{N-1}\right \rangle.
\end{equation}

Then, using the CNOT gate on the first two qubits, we can obtain
\begin{equation}
  \alpha_{N-1}\left |1\right \rangle \left |0\right \rangle^{\otimes N-1}+\left |0\right \rangle \left |\psi_O^{N-1}\right \rangle=\left |\psi_O^{N}\right \rangle.
\end{equation}

\subsection{$U_B^{(N)}$: From $\left | \psi_E^{(N)} \right \rangle$ to $\left | \psi_B^{(N)} \right \rangle$}

Starting with a simple case, if $N$ is 3, we can use a CNOT gate to transform the Edick state to the binary encoding state, which is the same as the one-hot encoding state.
\begin{equation}
  \alpha_0 \left | 00 \right \rangle+ \alpha_1 \left | 01 \right \rangle+\alpha_2 \left | 11 \right \rangle \rightarrow \alpha_0 \left | 00 \right \rangle+ \alpha_1 \left | 01 \right \rangle+\alpha_2 \left | 10 \right \rangle.
\end{equation}

 If $N$ is a small number ($N \leq 6$), the recursion method \cite{plesch_efficient_2010} introduced in Appendix \ref{a4} can prepare $\left | \psi_B^{(N)} \right \rangle$ from $\left | \psi_E^{(N)} \right \rangle$. However, the recursion method has $O(N\log N)$ depth, so it is not efficient when $N$ is large.

If $N$ is odd, we can devide $N-1$ qubits into two parts and use the divide-and-conquer method to prepare $ U_B^{N} $ from $ U_B^{(N+1)/2} $. The circuit is shown as Fig. \ref{f20} and a detailed example of $U_B^{(N)}$ is shown in Appendix \ref{a2}. The step of the scheme is as belows:

\begin{figure}[hbtp]
  \centering
  \includegraphics[width=0.5\textwidth]{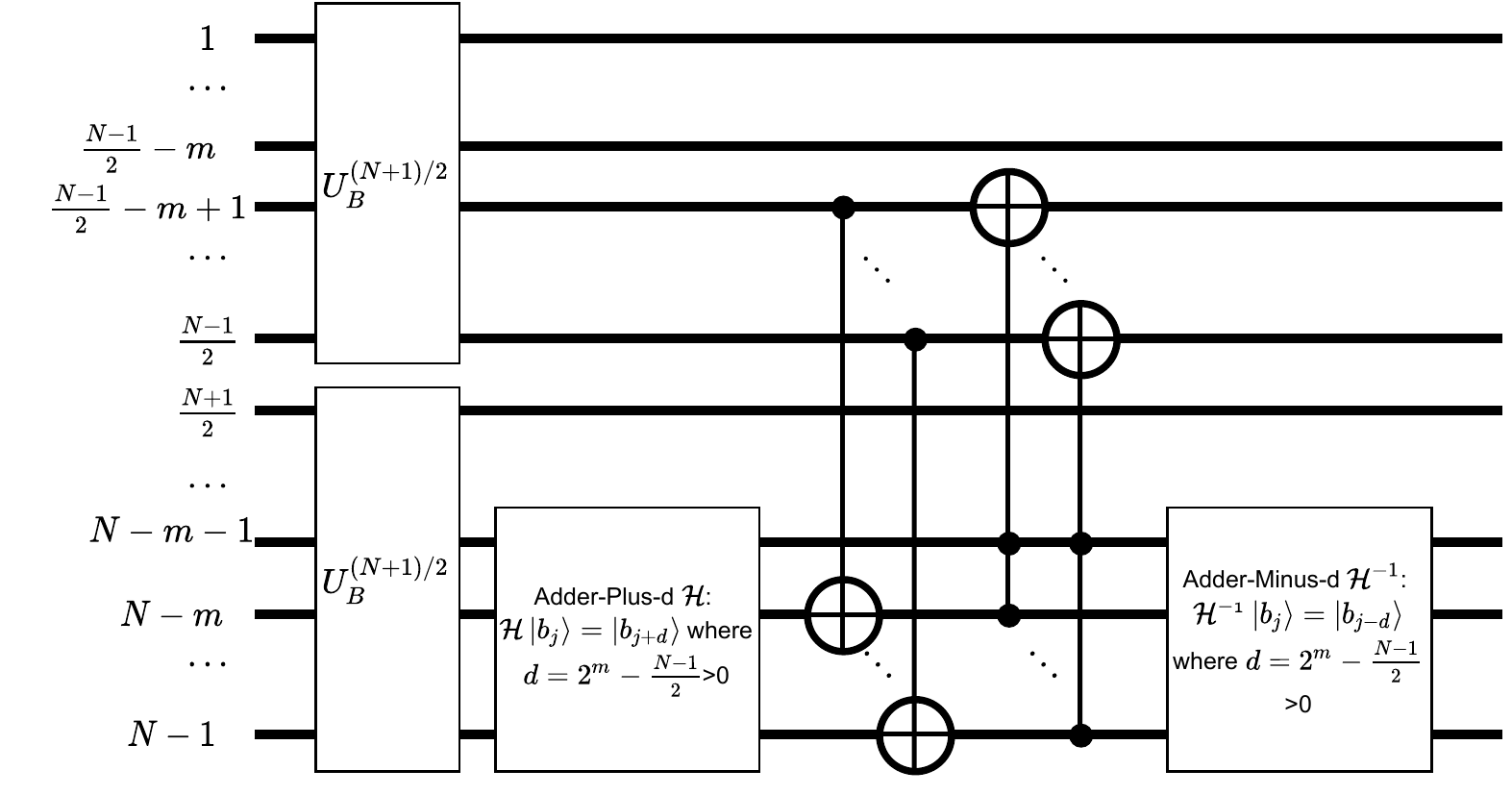}
  \caption{Edick to binary encoding state circuit $U_B^{(N)}$.}
  \label{f20}
\end{figure}

\textit{Step 0.} The Edick state can be rewritten as:

\begin{equation}
  \begin{aligned}
  &\sum_{i=0}^{(N-1)/2-1} \alpha_i \left |0\right \rangle^{\otimes \frac{N-1}{2}}(\left |0\right \rangle^{\otimes \frac{N-1}{2}-i}\left |1\right \rangle^{\otimes i})+\\
  &\sum_{i=(N-1)/2}^{N-1} \alpha_i (\left |0\right \rangle^{\otimes N-i-1}\left |1\right \rangle^{\otimes i-\frac{N-1}{2}})\left |1\right \rangle^{\otimes \frac{N-1}{2}}.
 \end{aligned}
\end{equation}

\textit{Step 1.} Implement $U_B^{(\frac{N-1}{2})}$ on the first part and the second part respectively.

\begin{equation}
  \begin{aligned}
  & \xrightarrow{U_B^{(\frac{N-1}{2})} \otimes U_B^{(\frac{N-1}{2})}}\sum_{i=0}^{(N-1)/2-1} \alpha_i \left |0\right \rangle^{\otimes \frac{N-1}{2}}(\left |0\right \rangle^{\otimes \frac{N-1}{2}-m}\left |b_i\right \rangle)\\
  &+\sum_{i=(N-1)/2}^{N-1} \alpha_i (\left |0\right \rangle^{\otimes \frac{N-1}{2}-m}\left |b_{i-(N-1)/2}\right\rangle)\\
  &(\left |0\right \rangle^{\otimes \frac{N-1}{2}-m} \left |b_{(N-1)/2}\right \rangle),
  \end{aligned}
\end{equation}
where $m=\lceil \log \frac{N-1}{2}\rceil$.

\textit{Step 2.} Apply adder-plus-$d$ gate $\mathcal{H}: \mathcal{H}\left |b_j \right \rangle=\left |b_{j+d} \right \rangle$ on the last $m+1$ qubit , where $d=2^m-\frac{N-1}{2}$. So if $d>0$, we have

\begin{align}
  \begin{aligned}
  &\xrightarrow{I \otimes \mathcal{H}}\sum_{i=0}^{(N-1)/2-1} \alpha_{i} \left |0\right \rangle^{\otimes \frac{N-1}{2}}(\left |0\right \rangle^{\otimes \frac{N-1}{2}-m}\left |b_{i+d}\right \rangle)\\
  &+\sum_{i=(N-1)/2}^{N-1} \alpha_i (\left |0\right \rangle^{\otimes \frac{N-1}{2}-m}\left |b_{i-(N-1)/2}\right\rangle)\\
  &(\left |0\right \rangle^{\otimes \frac{N-1}{2}-m-1} \left |1\right \rangle\left |0\right \rangle^{\otimes m}),
  \end{aligned}
\end{align}
where $\left |1\right \rangle\left |0\right \rangle^{\otimes m}=\left |b_{2^m}\right \rangle$. A scheme to implement the adder-plus-$d$ gate is shown in Appendix \ref{a3}.

\textit{Step 3.} For all $j \in \left [ \frac{N-1}{2}-m+1, \frac{N-1}{2} \right ]$, apply the CNOT gate, whose control qubit is $j$ th qubit and target qubit is $\frac{N+1}{2}+j$ th qubit.

\begin{align}
  \begin{aligned}
  &\xrightarrow{I \otimes CNOT} \sum_{i=0}^{(N-1)/2-1} \alpha_{i} \left |0\right \rangle^{\otimes \frac{N-1}{2}}(\left |0\right \rangle^{\otimes \frac{N-1}{2}-m}\left |b_{i+d}\right \rangle)\\
  &+\sum_{i=(N-1)/2}^{N-1} \alpha_i (\left |0\right \rangle^{\otimes \frac{N-1}{2}-m}\left |b_{i-(N-1)/2}\right\rangle)\\
  &(\left |0\right \rangle^{\otimes \frac{N-1}{2}-m-1} \left |1\right \rangle\left |b_{i-(N-1)/2}\right\rangle),
  \end{aligned}.
\end{align}

\textit{Step 4.} For all $j \in \left [ \frac{N-1}{2}-m+1, \frac{N-1}{2} \right ]$, apply the Toffoli gate, whose control qubits are $N-m-1$ th and $\frac{N-1}{2}+j$ th qubit and target qubit is $j$ th qubit.

\begin{align}
  \begin{aligned}
  &\xrightarrow{\text{Toffoli}} \sum_{i=0}^{(N-1)/2-1} \alpha_{i} \left |0\right \rangle^{\otimes \frac{N-1}{2}}(\left |0\right \rangle^{\otimes \frac{N-1}{2}-m}\left |b_{i+d}\right \rangle)\\
  &+\sum_{i=(N-1)/2}^{N-1} \alpha_i (\left |0\right \rangle^{\otimes \frac{N-1}{2}})
  (\left |0\right \rangle^{\otimes \frac{N-1}{2}-m-1} \\
  &\left |1\right \rangle\left |b_{i-(N-1)/2}\right\rangle)=\sum_{i=0}^{N-1} \alpha_i \left |0\right \rangle^{\otimes N-m-2} \left |b_{i+d}\right\rangle.
  \end{aligned}
\end{align}

\textit{Step 5.} Apply the inverse of the adder-plus-$d$ gate $\mathcal{H}^{-1}$ (adder-minus-$d$ gate) and we get the final state.
\begin{align}
  \begin{aligned}
  \xrightarrow{I \otimes \mathcal{H}^{-1}} \sum_{i=0}^{N-1} \alpha_i \left |0\right \rangle^{\otimes N-m-2} \left |b_{i}\right\rangle.
  \end{aligned}
\end{align}

\begin{figure}[hbtp]
  \centering
  \includegraphics[width=0.5\textwidth,trim=10 0 0 0, clip]{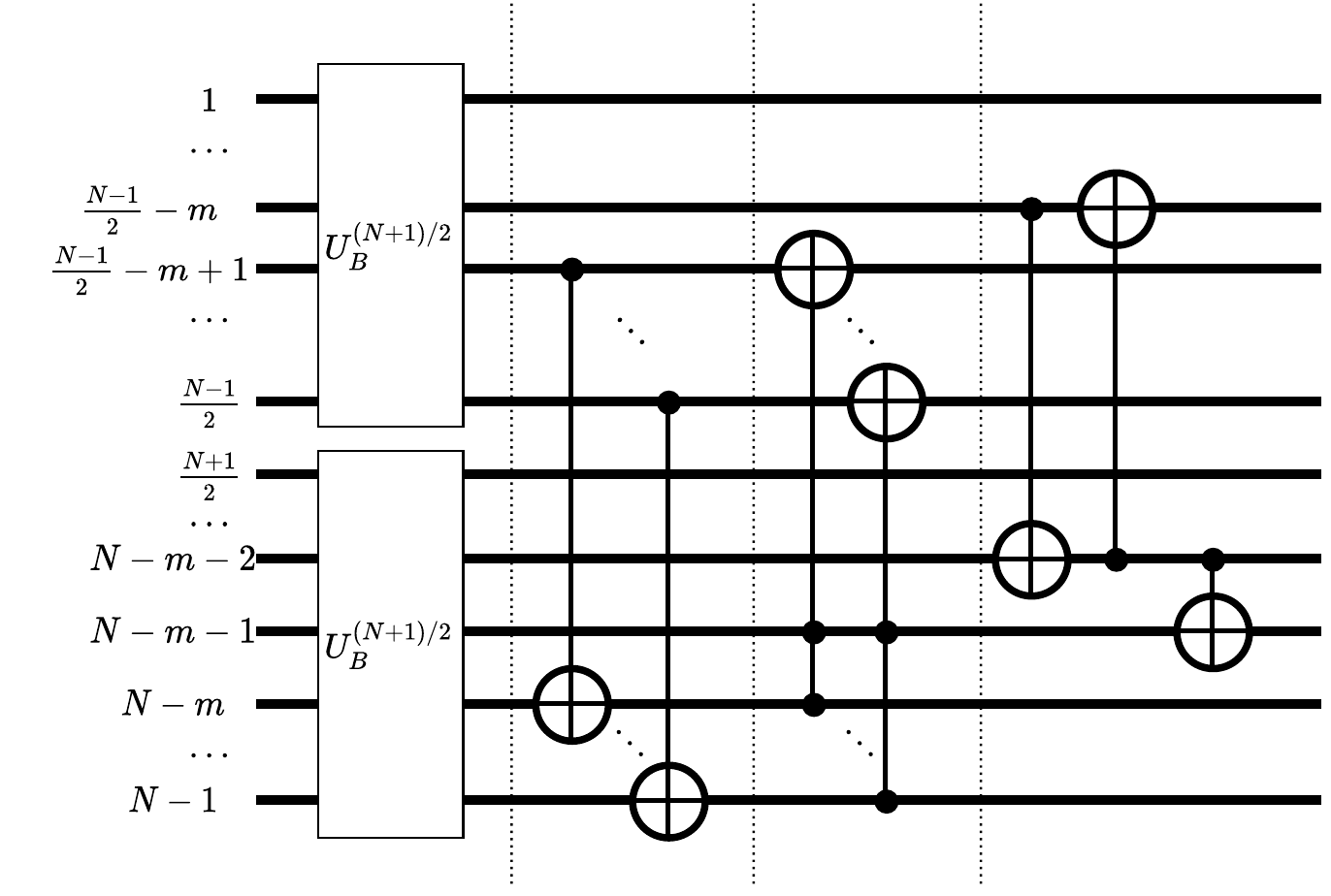}
  \caption{$U_B^{(N)}$ when $d=0$.}
  \label{f21}
\end{figure}

If $d=0$, \textit{step2} and \textit{step5} are canceled. After \textit{step1}, \textit{step3}, \textit{step4} the quantum state becomes:

\begin{align}
  \begin{aligned}
  &\sum_{i=0}^{(N-1)/2-1} \alpha_{i} \left |0\right \rangle^{\otimes \frac{N-1}{2}}(\left |0\right \rangle^{\otimes \frac{N-1}{2}-m}\left |b_{i}\right \rangle)+\\
  &\sum_{i=(N-1)/2}^{N-2} \alpha_i (\left |0\right \rangle^{\otimes \frac{N-1}{2}})(\left |0\right \rangle^{\otimes \frac{N-1}{2}-m-1} \left |1\right \rangle\left |b_{i-2^m}\right\rangle)+\\
  &\alpha_{N-1} (\left |0\right \rangle^{\otimes \frac{N-1}{2}-m-1} \left |1\right \rangle\left |0\right\rangle^{\otimes m})(\left |0\right \rangle^{\otimes \frac{N-1}{2}-m-1} \left |1\right \rangle\left |0\right\rangle^{\otimes m}).
  \label{e10}
  \end{aligned}
\end{align}

Then, we need an extra step with three CNOT gates as Fig. \ref{f21} shown. The contol qubits of the three CNOT gates are respectively  $\frac{N-1}{2}-m$ th, $N-m-2$ th and $N-m-2 th$ qubits and the target qubits are $N-m-2$ th,  $\frac{N-1}{2}-m$ th and $N-m-1$ th qubits. This action transforms

\begin{align}
  \begin{aligned}
    &\alpha_{N-1} (\left |0\right \rangle^{\otimes \frac{N-1}{2}-m-1} \left |1\right \rangle\left |0\right\rangle^{\otimes m})(\left |0\right \rangle^{\otimes \frac{N-1}{2}-m-1} \left |1\right \rangle\left |0\right\rangle^{\otimes m})\\&\xrightarrow{CNOT}  \alpha_{N-1}(\left |0\right \rangle^{\otimes \frac{N-1}{2}})(\left |0\right \rangle^{\otimes \frac{N-1}{2}-m-2}\left |1\right \rangle \left |0\right \rangle ^{\otimes m+1})
  \end{aligned}
\end{align}.

and keep the other terms in the Eq. \ref{e10} unchanged. Therefore, the final state becomes
\begin{align}
  \begin{aligned}
  \sum_{i=0}^{N-1} \alpha_i \left |0\right \rangle^{\otimes N-m-3} \left |b_{i}\right\rangle.
  \end{aligned}
\end{align}

If $N$ is even, three methods with different depth and ancilla number (seen in Sec. \ref{s1} and Sec. \ref{sec4}) are used to prepare $\left | U_B^{(N)} \right \rangle$. These three methods are: (\uppercase\expandafter{\romannumeral1}). the \textit{ recursion} method, (\uppercase\expandafter{\romannumeral2}). the \textit{expanding-to-$N+1$} method, and (\uppercase\expandafter{\romannumeral3}). the \textit{ expanding-to-$2^k$} method.

The \textit{ recursion} method is shown in Appendix \ref{a4} and uses $U_B^{(N-1)}$ to prepare $U_B^{(N)}$ without ancilla. The \textit{expanding-to-$N+1$} method is to prepare $U_B^{(N+1)}$ to replace $U_B^{(N)}$ and the \textit{ expanding-to-$2^k$} is to prepare $U_B^{(2^k+1)}$ to replace $U_B^{(N)}$. This is because for all $i>j$, $U_B^{(i)}$ can also transform $\left | \psi_E^{(j)} \right \rangle$ to $\left | \psi_B^{(j)} \right \rangle$.

\subsection{Circuit depth and size}
\label{s1}
From Fig. \ref{f9},  $d_o(N)$, the depth of $U_O^{(N)}$ satisfies
\begin{align}
  \begin{aligned}
    d_o(N)&=d_o(N/2)+2, \text{if $N$ is even}\\
    d_o(N)&=d_o(N-1), \text{if $N$ is odd}\\
    d_o(2)&=1.
    \label{e02}
  \end{aligned}
\end{align}

and $s_o(N)$, the size of $U_O^{(N)}$ satisfies
\begin{align}
  \begin{aligned}
    s_o(N)&=s(N/2)+N-1, \text{if $N$ is even}\\
    s_o(N)&=s(N-1)+1, \text{if $N$ is odd}\\
    s_o(2)&=1.
    \label{e03}
  \end{aligned}
\end{align}

The reason of the second equation in Eq. \ref{e02} is that the CNOT gate in Fig. \ref{f8} can be combined with the last $(N-1)/2-1$ gates in Fig. \ref{f9}. From Eq. \ref{e02} and \ref{e03}, we have $d_o(N)=2\lceil \log N \rceil-1$ and  $s_o(N)<1+N+\log N$ (proof in Appendix \ref{a6} ) Therefore, $U_O^{(N)}$'s depth is $O(\log N)$ and size is $O(N)$.

As to $U_B^{(N)}$,  if we use the \textit{expanding-to-$2^k$} method, we only calculate the complexity when $N=2^{m+1}+1$, which satisfies:
\begin{align}
  \begin{aligned}
    d_b(N)&=d_b(\frac{N+1}{2})+O(\log N),\\
    s_b(N)&=2s_b(\frac{N+1}{2})+O(\log N). \\
    \label{e04}
  \end{aligned}
\end{align}

If we use the \textit{expanding-to-$N+1$} method, according to Fig. \ref{f20}, the adder-plus-$d$ gate $\mathcal{H}$ can be implemented by $O(m)=O(\log N)$ depth and $O(m^2)=O(\log^2 N)$ size (the Fourier method in Appendix \ref{a3}). Moreover, CNOT gates' depth is $O(1)$ and size is $O(m)$, and Toffoli gates' depth is $O(m)$ and size is $O(m)$. By the way, if we replace the Fourier adder-plus-$d$ gate with the carry-lookahead adder (QLCA) gate \cite{draper_logarithmic-depth_2006}, the depth of the adder gate drops to $O(\log m)$. However, the depth of the entire circuit remains the same.

Therefore, it is clear that

\begin{align}
  \begin{aligned}
    d_b(N)&=d_b(\frac{N+1}{2})+O(\log N), \text{if $N$ is odd.}\\
    d_b(N)&=d_b(N+1), \text{if $N$ is even.} \\
    \label{e07}
  \end{aligned}
\end{align}
and
\begin{align}
  \begin{aligned}
    s_b(N)&=2s_b(\frac{N+1}{2})+O(\log^2 N), \text{if $N$ is odd.}\\
    s_b(N)&=s_b(N+1), \text{if $N$ is even.} \\
    \label{e08}
  \end{aligned}
\end{align}

If we use the \textit{recursion} method,  CNOT gates ($\log N$ depth and $\log N$ size) and  multiple qubit Toffoli gates($O(\log N)$ depth and $O(\log^2 N)$ size \cite{saeedi_linear-depth_2013}) should be added to the circuit. Therefore, the whole circuit satisfies that
\begin{align}
  \begin{aligned}
    d_b(N)&=d_b(\frac{N+1}{2})+O(\log N), \text{if $N$ is odd.}\\
    d_b(N)&=d_b(N+1)+O(\log N), \text{if $N$ is even.} \\
    \label{e05}
  \end{aligned}
\end{align}
and
\begin{align}
  \begin{aligned}
    s_b(N)&=2s_b(\frac{N+1}{2})+O(\log^2 N), \text{if $N$ is odd.}\\
    s_b(N)&=s_b(N-1)+O(\log^2 N), \text{if $N$ is even.} \\
    \label{e06}
  \end{aligned}
\end{align}

After solving Eq. \ref{e04} $\sim$ Eq. \ref{e06} (proof in Appendix \ref{a6}), we come to the conclusion that $d_b(N)=O(\log^2 N)$ and $s_b(N)=O(N)$ no matter which method we choose among (\uppercase\expandafter{\romannumeral1}). the \textit{recursion} method,  (\uppercase\expandafter{\romannumeral2}). the \textit{ expanding-to-$N+1$} method or (\uppercase\expandafter{\romannumeral3}). the \textit{expanding-to-$2^k$} method.

After combining $U_O^{(N)}$ and $U_B^{(N)}$, the entire circuit has $O(\log^2 N)$ depth and $O(N)$ size.

\section{Preparation of the binomial distribution state}
\label{sec3}

By using the idea in \cite{bartschi_deterministic_2019}, we can use RY gate to prepare the binomial Edick state $\left | \psi_E^{(N+1)} \right \rangle=\sum_{i=0}^{N} \sqrt{f(i, p)} \left |0 \right \rangle^{\otimes N-i} \left |1 \right \rangle ^{\otimes i}$ where $f(i, p) = \binom {N}{i} p^i (1-p)^{N-i}$ is the probability density function of the binomial distribution $\mathcal{B}(N, p)$. We can also simulate the normal distribution in this way, because according to the central limit theorem, the binomial distribution $\mathcal{B}(n, p)$ tends to the normal distribution $\mathcal{N}(np, np(1-p))$ as $n$ increases. Then, we use the converter in Sec. \ref{sec2} to prepare the one-hot binomial state and the binary binomial state.  The circuit is shown as Fig. \ref{f5} and a detailed introduction is shown in Appendix \ref{a5}.

\begin{figure}[hbtp]
  \centering
  \includegraphics[width=0.5\textwidth,trim=50 0 0 0, clip]{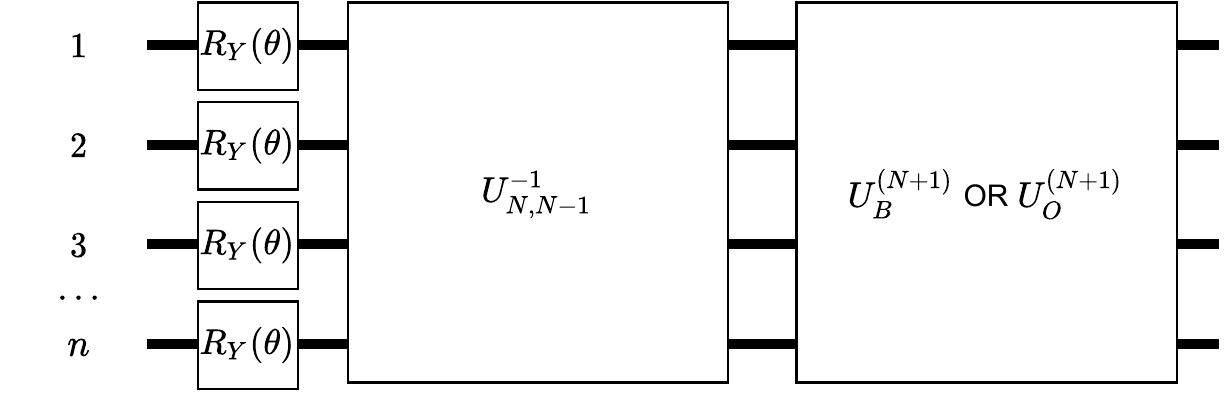}
  \caption{Generation of the binomial distribution state.}
  \label{f5}
\end{figure}

The reference \cite{bartschi_deterministic_2019} has proved that the depth and the size of $U_{N, N-1}$ are $O(N)$ and $O(N^2)$. Therefore, according to Sec. \ref{s1}, the whole binomial state generation circuit's depth is $O(N)$ and the size is $O(N^2)$. Moreover, due to the binomial theorem, our scheme has no error in theory when preparing binomial distribution states.

If we want to prepare a normal distribution, we can adjust the value of $\theta$ to determine the variance.  Since the variance
\begin{equation}
  \sigma = Np(1-p)=\frac{N\sin^2 \theta}{4},
\end{equation}
an N-qubit circuit can simulate any variance from 0 to $\frac{N}{4}$ by only adjusting the parameter $\theta$.

\section{Comparation with other works}
\label{sec5}
Plesch $et$ $al$. \cite{plesch_efficient_2010} propose a method to implement the quantum compression, which is applied to prepare the binomial distribution state and then B$\ddot a$rtschi $et$ $al$. \cite{bartschi_deterministic_2019} solve this problem by the Dicke state. Both schemes use the converter scheme as a intermediate step. Plesch's method has $O(N \log N)$ depth and $O(N \log^2 N)$ size and B$\ddot a$rtschi's method improve the size to $O(N \log N)$. Xiaoming Sun $et$ $al$.\cite{sun_asymptotically_2021} propose a conversion method when $N=2^n$. With $2N$ ancilla qubits, the conversion can be implemented by a quantum circuit using single-qubit gates and CNOT gate of depth $O(\log N)$ and size $O(N)$. Compared to these schemes, our converter achieves an exponential speedup with $O(\log^2 N)$ depth and $O(N)$ size without ancilla.

\begin{figure}[hbtp]
  \centering
  \includegraphics[width=0.55\textwidth,trim=50 0 0 0, clip]{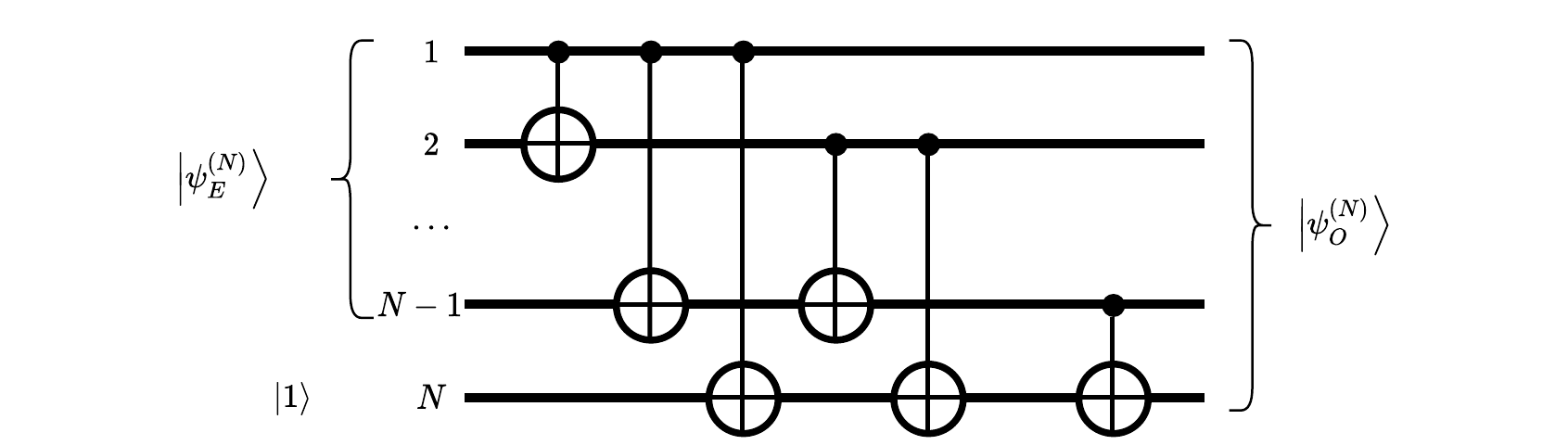}
  \caption{$O(N)$-depth Edick to one-hot encoding state circuit.}
  \label{f1}
\end{figure}

B$\ddot a$rtschi $et$ $al$.\cite{bartschi_deterministic_2019} also propose
a method called CNOT stair to generate one-hot encoding state from the Edick state circuit shown as Fig. \ref{f1}. The qubit  $\left |1 \right \rangle$ should be tensored with $\left | \psi_E^{(N)} \right \rangle$ first and then $U_O^{(N)}$ applies $O(N^2)$ CNOT gates on all $N$ qubits. This circuit includes $\frac{1}{2} N^2$ CNOT gates and its depth is $2*N-3$. In contrast, our scheme can achieve exponentially speed up in this part by $O(\log N)$ depth and $O(N)$ size.

Under the support of the converter scheme, the complexity of preparing binomial distribution states also decreases slightly compared to the previous two schemes \cite{plesch_efficient_2010, bartschi_deterministic_2019}. The depth of Plesch's method and B$\ddot a$rtschi's method are $O(N^2)$ and $O(N\log N)$ respectively, while our method's depth is $O(N)$. Compared with the arbitrary preparation methods \cite{grover_creating_2002, mottonen_transformation_2005, iten_quantum_2016} that has the same complexity with our scheme, we do not need to calculate the value of $f(k, p)$ in classcial way. The classical complexity to calculate $f(k, p)$ should be more than $O(N)$, because as $N$ increases, the computational complexity of a single probability increases, and the computation needs to be done $N+1$ times from $f(0, p)$ to $f(N, p)$. Besides, from a $\mathcal{B}(n,p)$-state preparation to a $\mathcal{B}(n,q)$-state preparation, only 1-depth circuit need to be changed (only change the parameter of the first $N$ $R_y$ gates), but the arbitrary preparation methods need to change massive rotation gates.

A recent paper \cite{rattew_efficient_2021} using mid-circuit partial measurements has shown that an $O(\log N)$-depth circuit has ability to prepare the normal distribution state with the success probability of more than $\frac{1}{2^{t_1}}$, where $t_1$ is a constant variable decides the qubits counts and the circuit depth. This scheme also uses the central limit theorem to prepare the normal distribution by quantum random walks. In contrast, our circuit does not require the manipulation of intermediate measurements so it is deterministic and is easier to prepare. Moreover, when the variance of the normal distribution is within a certain range ($0 \sim \frac{N}{4}$), we can realize the preparation of the quantum state only by changing the angle of the $R_y$ gate, while method in \cite{rattew_efficient_2021} should change the number of iterations of the circuit to approximate the target variance.

\section{Numerical simulation}
\label{sec4}

In this section, we decompose the converter $U_O (U_B^{-1}\otimes I)$ into single qubit gates and CNOT gates in Qiskit, and then count the circuit depth and size of three methods: (\uppercase\expandafter{\romannumeral1}). the  \textit{recursion} method, (\uppercase\expandafter{\romannumeral2}), the  \textit{expanding-to-$N+1$} method (\uppercase\expandafter{\romannumeral3}). the  \textit{expanding-to-$2^k$} method.

\begin{figure}[hbtp]
  \centering
  \includegraphics[width=0.5\textwidth,trim=50 0 0 0, clip]{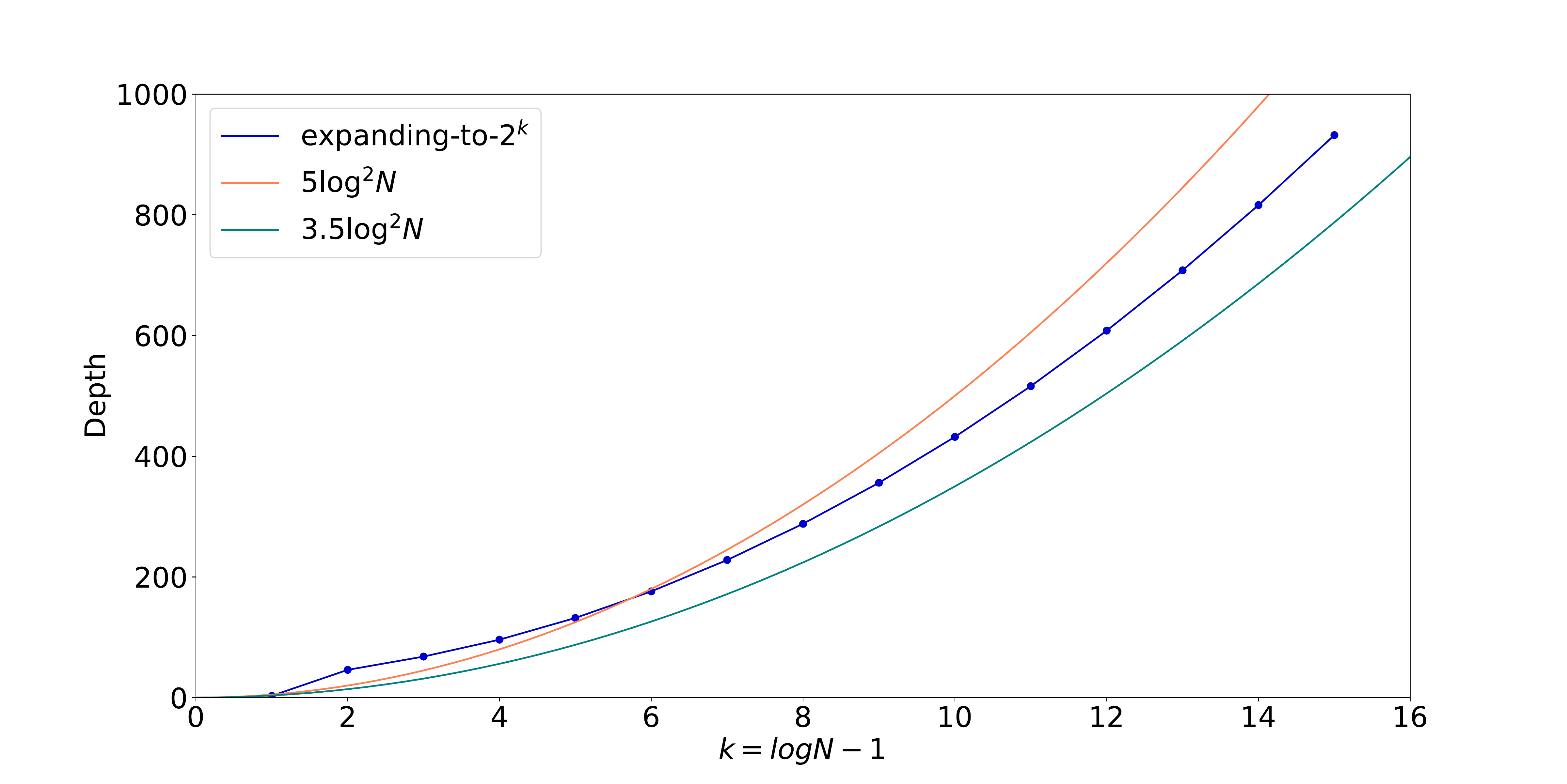}
  \caption{The depth of the converter circuit when $N=2^k+1$.}
  \label{f10}
\end{figure}

\begin{figure}[hbtp]
  \centering
  \includegraphics[width=0.5\textwidth,trim=50 0 0 0, clip]{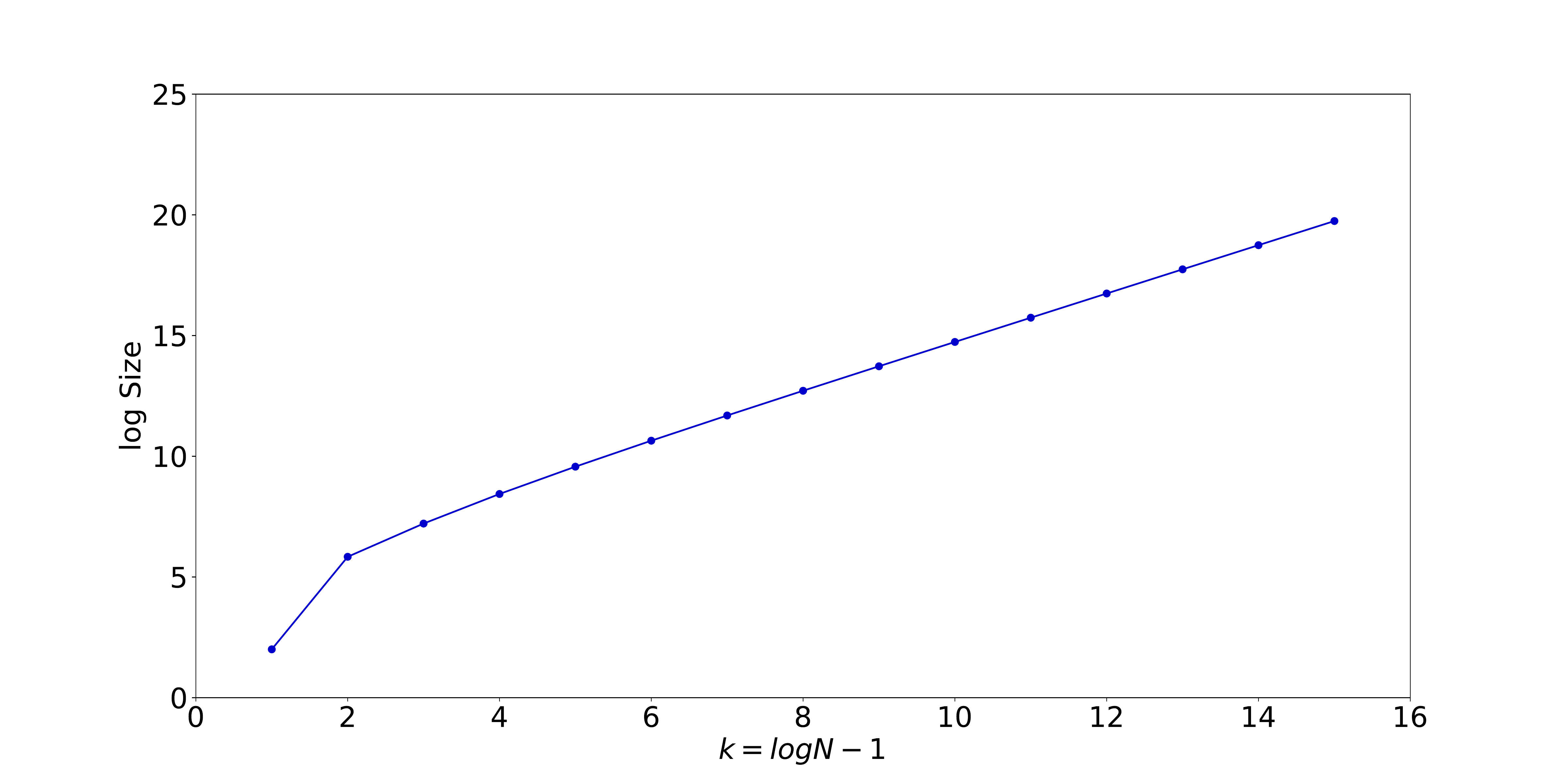}
  \caption{The size of the converter circuit when $N=2^k+1$.}
  \label{f11}
\end{figure}

Given that $N=2^k+1$ and $k$ from 1 to 15,  Fig. \ref{f10} shows the converter depth is near $4 k^2$ and Fig. \ref{f11} shows the size is linear with $N$, which is approximate to $23N$. Therefore, if we use the \textit{Expanding-to-$2^k$} method, the depth is between $4 \log^2 N$ to $4 \log^2 2N$ and the size is between $23N$ to $46N$.

\begin{figure}[hbtp]
  \centering
  \includegraphics[width=0.5\textwidth,trim=50 0 0 0, clip]{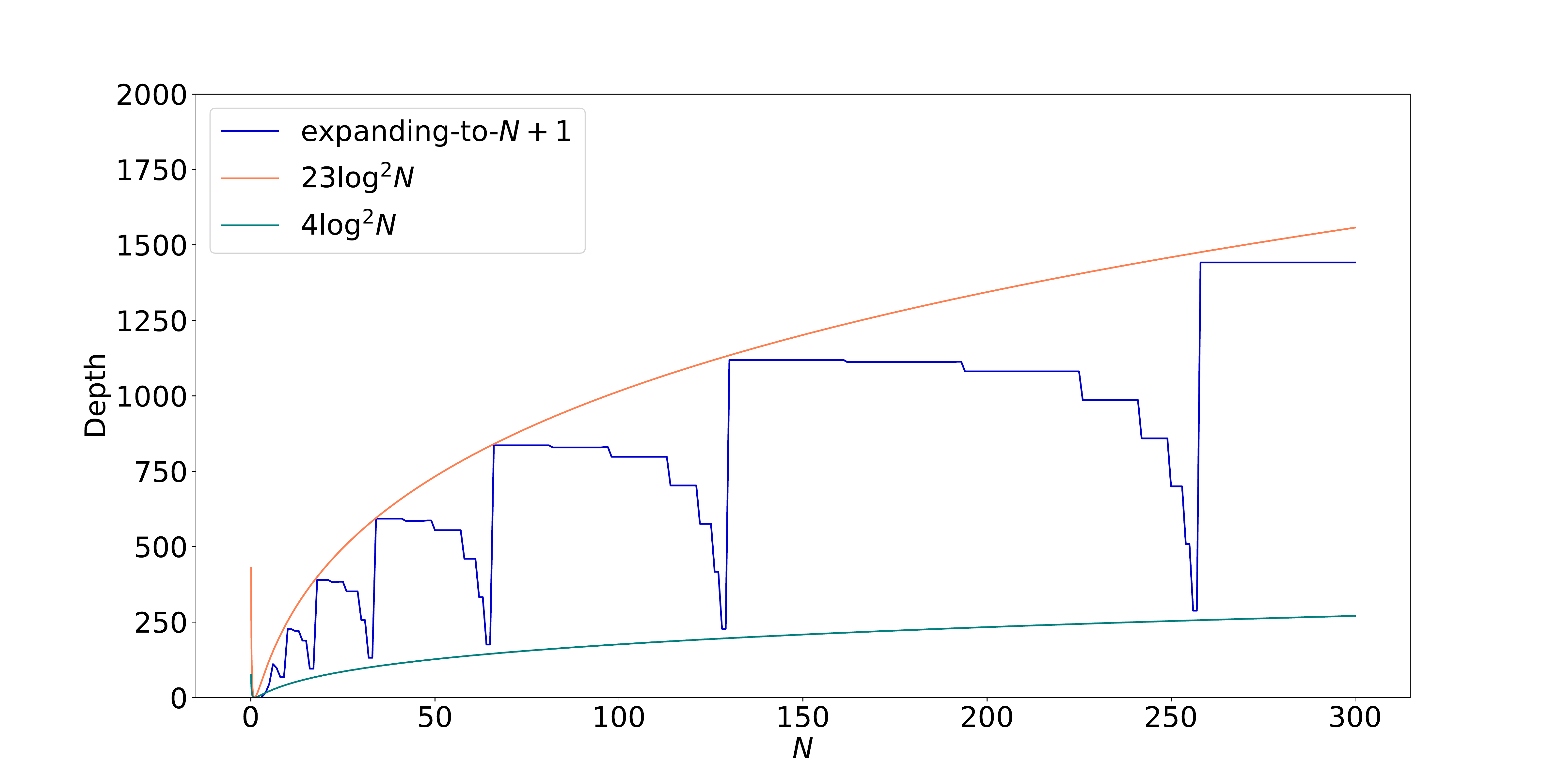}
  \caption{The depth of the converter circuit using \textit{Expanding-to-$N+1$}.}
  \label{f12}
\end{figure}

\begin{figure}[hbtp]
  \centering
  \includegraphics[width=0.5\textwidth,trim=50 0 0 0, clip]{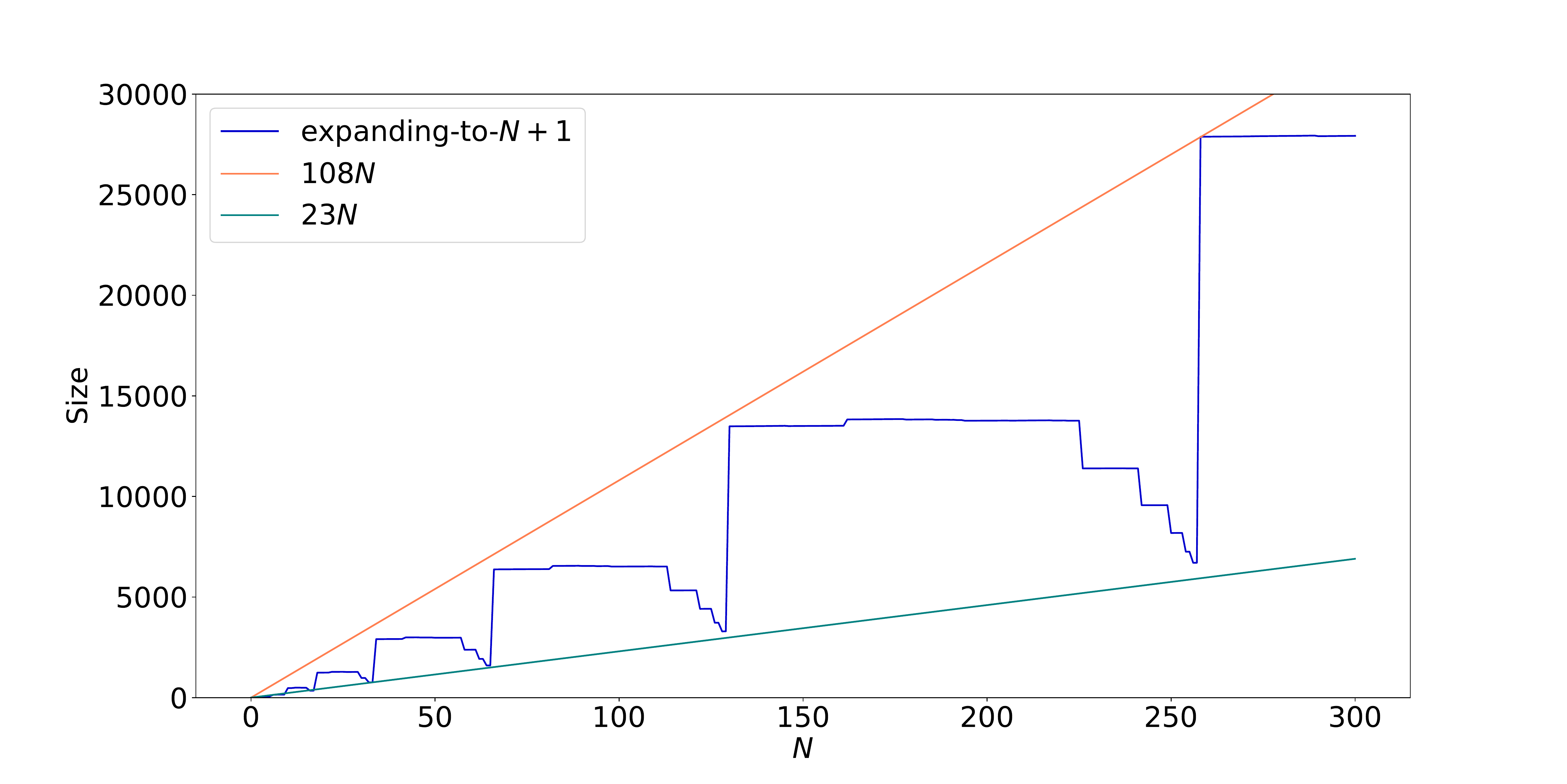}
  \caption{The size of the converter circuit using \textit{Expanding-to-$N+1$}.}
  \label{f13}
\end{figure}

\begin{figure}[hbtp]
  \centering
  \includegraphics[width=0.5\textwidth,trim=50 0 0 0, clip]{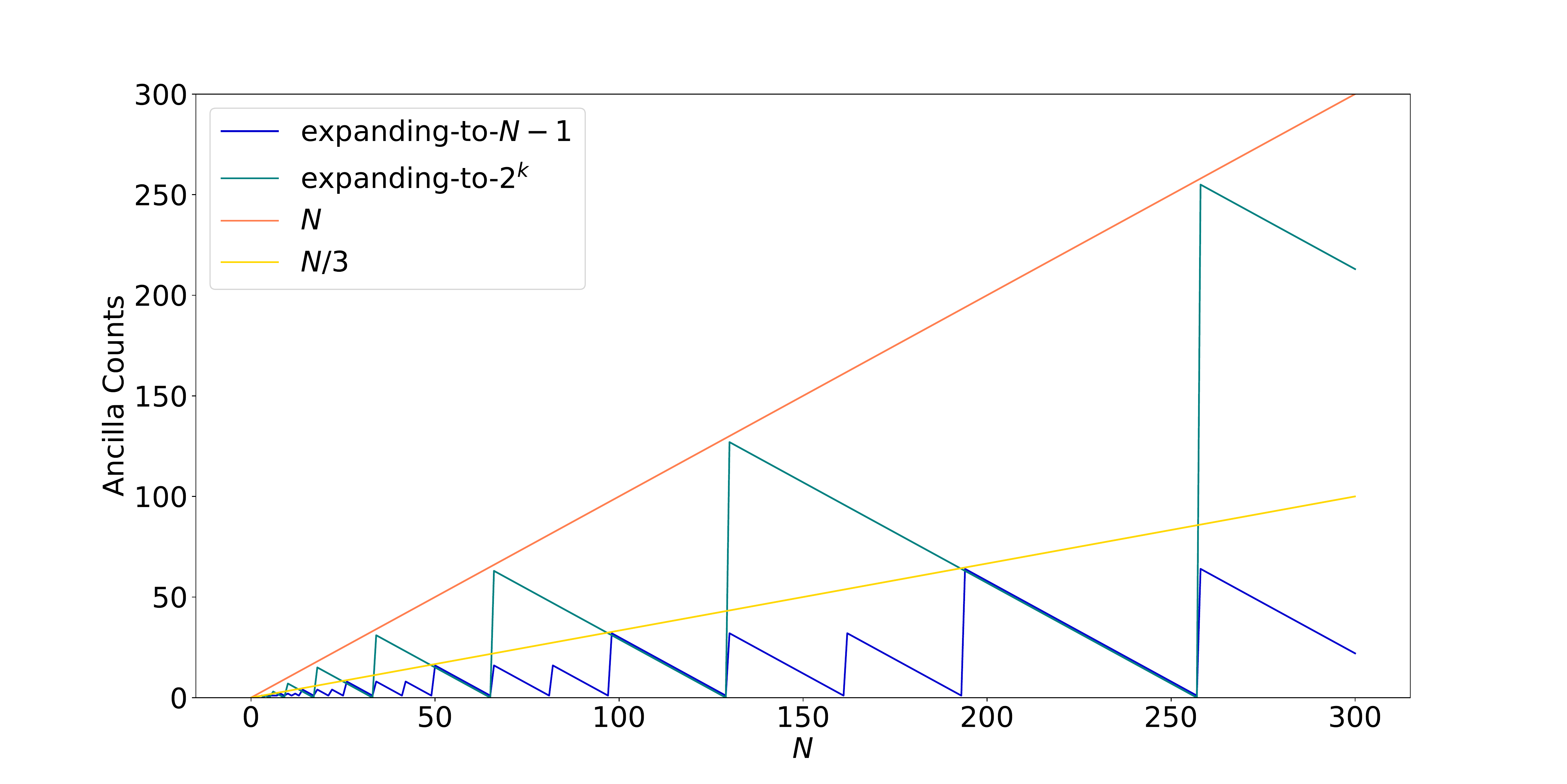}
  \caption{The number of ancilla in \textit{Expanding-to-$N+1$} and \textit{Expanding-to-$2^k$}.}
  \label{f14}
\end{figure}

Fig. \ref{f12} and Fig. \ref{f13} show the depth and size of the circuit by using \textit{expanding-to-$N+1$} method when $N$ is from 3 to 300. The depth is between $4 \log^2 N$ and $23 \log^2N$ and the size is between $23 N$ to $108 N$. As $N$ approaches the integer power of 2, the lines in two figures drop with a decreasing step function and jump suddenly because the number of adder-plus-$d$ gates is reduced. The number of the ancilla in \textit{expanding-to-$2^k$} and \textit{expanding-to-$N+1$} is shown as Fig. \ref{f14}. The \textit{expanding-to-$2^k$} method requires at most $N-1$ qubits while \textit{expanding-to-$N+1$} requires at most $\frac{N-5}{3}$ qubits.

\begin{figure}[hbtp]
  \centering
  \includegraphics[width=0.5\textwidth,trim=50 0 0 0, clip]{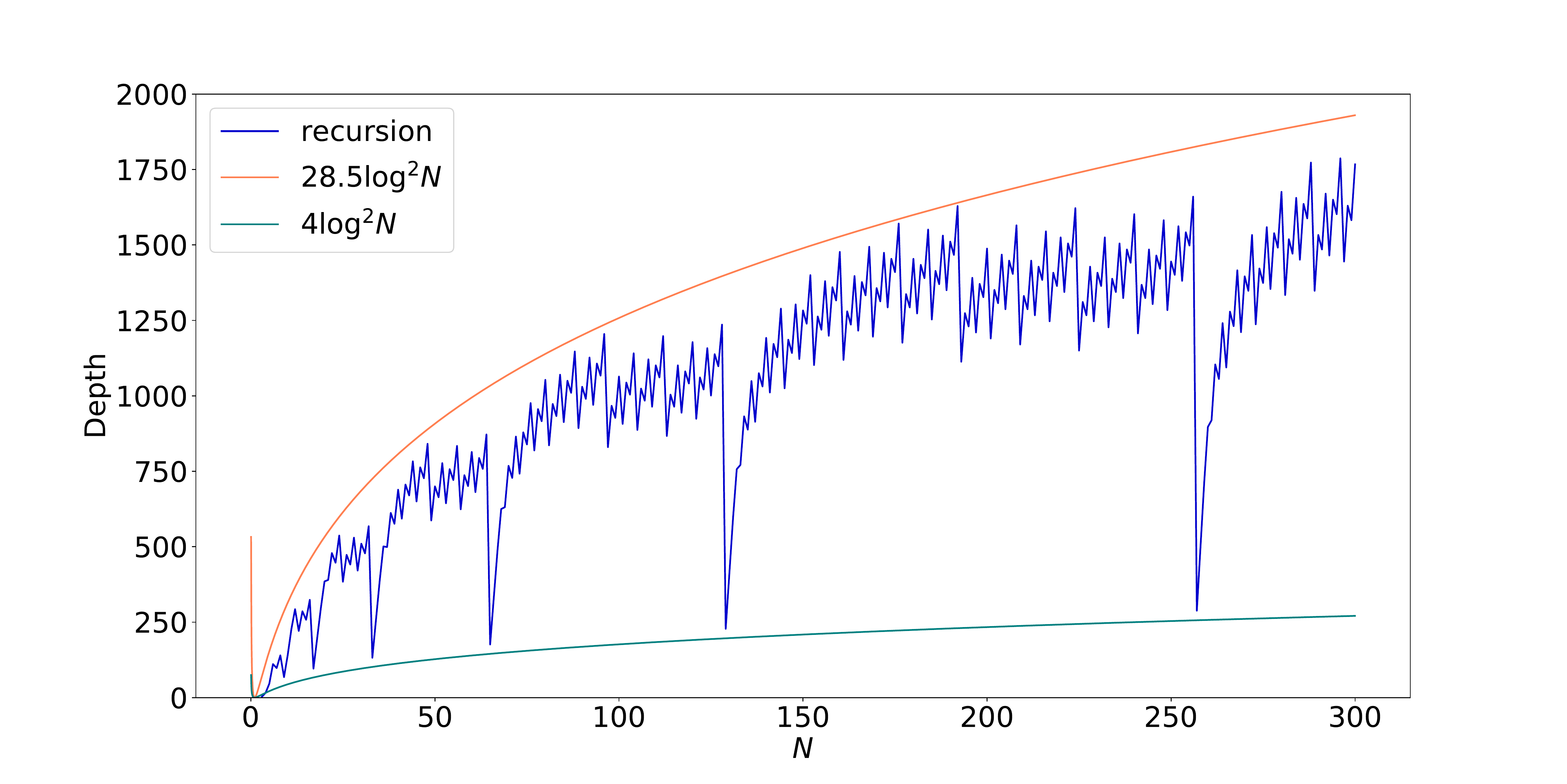}
  \caption{The depth of the converter circuit using \textit{recursion}.}
  \label{f15}
\end{figure}

\begin{figure}[hbtp]
  \centering
  \includegraphics[width=0.5\textwidth,trim=50 0 0 0, clip]{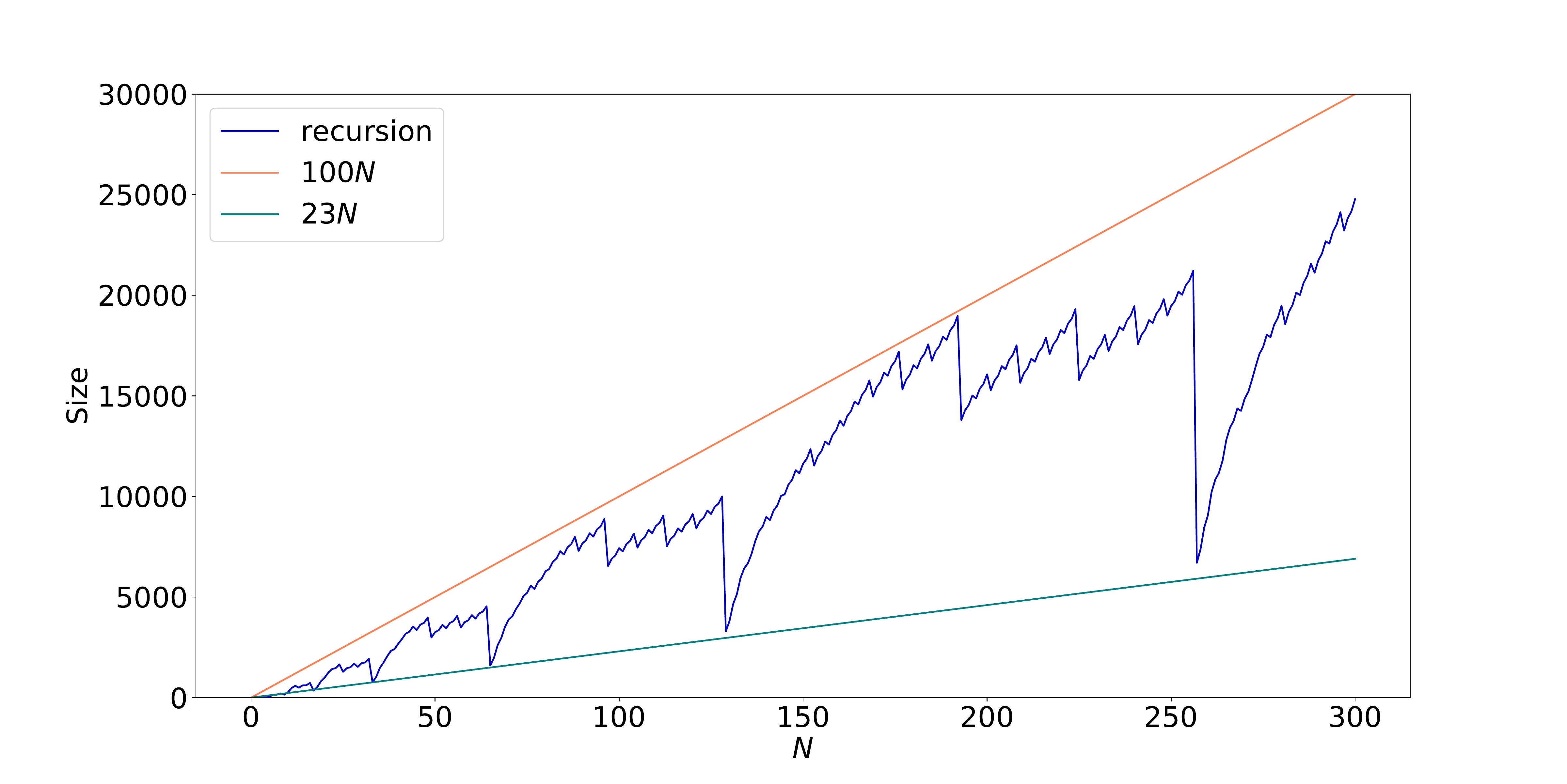}
  \caption{The size of the converter circuit using \textit{recursion}.}
  \label{f17}
\end{figure}

Fig. \ref{f15} and Fig. \ref{f17} show the depth and size of the circuit by using \textit{recursion} method when $N$ is from 3 to 300. The depth is between $4 \log^2 N$ and $28.5 \log^2N$ and the size is between $23 N$ to $100 N$. Without ancilla, the constant of the \textit{recursion} is approximate to that of the \textit{Expanding-to-$N+1$} method. Different from Fig. \ref{f12} and Fig. \ref{f13}, the lines in Fig. \ref{f14} and Fig. \ref{f15} drop sharply and rise gradually when $N$ is near the integer power of 2.


\section{Conclusion}
\label{sec6}

In conclusion, we presented a scheme to convert between the one-hot encoding state and the binary encoding state. The entire converter circuit consists of the Edick to one-hot circuit and the Edick to binary circuit. With the combination of the two parts, the circuit depth is $O(\log^2 N)$ and size is $O(N)$, which improves upon the previous work.

We also use this converter to prepare binomial distribution quantum states. Due to the central limit theorem, this quantum state can approximate the normal distribution state as $N$ increases. The state generation circuit has $O(N)$ depth and $O(N^2)$ size. If there is a better way to prepare the Edick state, the depth and size of the line have the potential to be further reduced.

In the simulation, we find the the lower bound of the converter depth is $4\log ^2 N$ and that of the converter size is $23 N$. If we use the \textit{expanding-to-$2^k$} method, by adding at most $N$ ancilla, the upper bound of the depth is $4\log^2 2N$ and that of the size is $46N$. If we use the \textit{expanding-to-$N+1$} method, by adding at most $N/3$ ancilla, the upper bound of the depth is $23\log ^2N$ and that of the size is $108N$. If we use the \textit{recursion} method, without ancilla, the upper bound of the depth is $28.5\log ^2N$ and that of the size is $100N$.

\bibliographystyle{unsrt}  
\bibliography{references}  

\begin{thebibliography}{10}

\bibitem{farhi_quantum_2014}
Edward Farhi, Jeffrey Goldstone, and Sam Gutmann.
\newblock A quantum approximate optimization algorithm.

\bibitem{hadfield_quantum_2019}
Stuart Hadfield, Zhihui Wang, Bryan O'Gorman, Eleanor~G. Rieffel, Davide
  Venturelli, and Rupak Biswas.
\newblock From the quantum approximate optimization algorithm to a quantum
  alternating operator ansatz.
\newblock 12(2):34.

\bibitem{harrow_quantum_2009}
Aram~W. Harrow, Avinatan Hassidim, and Seth Lloyd.
\newblock Quantum algorithm for linear systems of equations.
\newblock 103(15):150502.

\bibitem{cong_quantum_2016}
Iris Cong and Luming Duan.
\newblock Quantum discriminant analysis for dimensionality reduction and
  classification.
\newblock 18(7):073011.

\bibitem{cruz_efficient_2019}
Diogo Cruz, Romain Fournier, Fabien Gremion, Alix Jeannerot, Kenichi Komagata,
  Tara Tosic, Jarla Thiesbrummel, Chun~Lam Chan, Nicolas Macris, Marc‐André
  Dupertuis, and Clément Javerzac‐Galy.
\newblock Efficient quantum algorithms for {GHZ} and \$w\$ states, and
  implementation on the {IBM} quantum computer.
\newblock 2(5):1900015.

\bibitem{rebentrost_quantum_2018}
Patrick Rebentrost, Brajesh Gupt, and Thomas~R. Bromley.
\newblock Quantum computational finance: Monte carlo pricing of financial
  derivatives.
\newblock 98(2):022321.

\bibitem{grover_creating_2002}
Lov Grover and Terry Rudolph.
\newblock Creating superpositions that correspond to efficiently integrable
  probability distributions.

\bibitem{mottonen_transformation_2005}
Mikko Möttönen, Juha~J. Vartiainen, Ville Bergholm, and Martti~M. Salomaa.
\newblock Transformation of quantum states using uniformly controlled
  rotations.
\newblock 5(6):467--473.

\bibitem{iten_quantum_2016}
Raban Iten, Roger Colbeck, Ivan Kukuljan, Jonathan Home, and Matthias
  Christandl.
\newblock Quantum circuits for isometries.
\newblock 93(3):032318.

\bibitem{plesch_efficient_2010}
Martin Plesch and Vladimir Buzek.
\newblock Efficient compression of quantum information.
\newblock 81(3):032317.

\bibitem{bartschi_deterministic_2019}
Andreas Bärtschi and Stephan~J. Eidenbenz.
\newblock Deterministic preparation of dicke states.
\newblock In Leszek~Antoni Gasieniec, Jesper Jansson, and Christos Levcopoulos,
  editors, {\em Fundamentals of Computation Theory - 22nd International
  Symposium, {FCT} 2019, Copenhagen, Denmark, August 12-14, 2019, Proceedings},
  volume 11651 of {\em Lecture Notes in Computer Science}, pages 126--139.
  Springer.

\bibitem{abdullah-al-shafi_new_2017}
Md~Abdullah-Al-Shafi and Ali~Newaz Bahar.
\newblock A new approach of presenting binary to grey and grey to binary code
  converter in majority voter-based {QCA} nanocomputing.
\newblock 14(5):2416--2421.

\bibitem{das_reversible_2015}
Jadav~Chandra Das and Debashis De.
\newblock Reversible binary to grey and grey to binary code converter using
  {QCA}.
\newblock 61(3):223--229.

\bibitem{karkaj_binary_2017}
Ehsan~Taher Karkaj and Saeed~Rasouli Heikalabad.
\newblock Binary to gray and gray to binary converter in quantum-dot cellular
  automata.
\newblock 130:981--989.

\bibitem{woerner_quantum_2019}
Stefan Woerner and Daniel~J. Egger.
\newblock Quantum risk analysis.
\newblock 5(1):15.

\bibitem{draper_logarithmic-depth_2006}
T.G. Draper, S.A. Kutin, E.M. Rains, and K.M. Svore.
\newblock A logarithmic-depth quantum carry-lookahead adder.
\newblock 6(4):351--369.

\bibitem{saeedi_linear-depth_2013}
Mehdi Saeedi and Massoud Pedram.
\newblock Linear-depth quantum circuits for n -qubit toffoli gates with no
  ancilla.
\newblock 87(6):062318.

\bibitem{sun_asymptotically_2021}
Xiaoming Sun, Guojing Tian, Shuai Yang, Pei Yuan, and Shengyu Zhang.
\newblock Asymptotically optimal circuit depth for quantum state preparation
  and general unitary synthesis.

\bibitem{rattew_efficient_2021}
Arthur~G. Rattew, Yue Sun, Pierre Minssen, and Marco Pistoia.
\newblock The efficient preparation of normal distributions in quantum
  registers.
\newblock 5:609.

\end{thebibliography}

\begin{appendices}
  \section{The procedure in $U_O^{(4)}$ and $U_O^{(5)}$ }
\label{a1}
Here, we give the transformation of the quantum state in $U_O^{(4)}$ and $U_O^{(5)}$.

$U_O^{(4)}$:
\begin{align*}
  \left |000\right \rangle \xrightarrow{\otimes \left |1\right \rangle } \left |0001\right \rangle \xrightarrow{CNOT \text{ 1 to 2, 3 to 4}} \left |0001\right \rangle \xrightarrow{U_O^{(2)} \text{ 1 to 3}} \left |0001\right \rangle \xrightarrow{CNOT \text{ 2 to 3}} \left |0001\right \rangle\\
  \left |001\right \rangle \xrightarrow{\otimes \left |1\right \rangle } \left |0011\right \rangle \xrightarrow{CNOT \text{ 1 to 2, 3 to 4}} \left |0010\right \rangle \xrightarrow{U_O^{(2)} \text{ 1 to 3}}\left |0010\right \rangle \xrightarrow{CNOT \text{ 2 to 3}} \left |0010\right \rangle\\
  \left |011\right \rangle \xrightarrow{\otimes \left |1\right \rangle } \left |0111\right \rangle \xrightarrow{CNOT \text{ 1 to 2, 3 to 4}} \left |0110\right \rangle \xrightarrow{U_O^{(2)} \text{ 1 to 3}}\left |0110\right \rangle \xrightarrow{CNOT \text{ 2 to 3}} \left |0100\right \rangle\\
  \left |111\right \rangle \xrightarrow{\otimes \left |1\right \rangle } \left |1111\right \rangle \xrightarrow{CNOT \text{ 1 to 2, 3 to 4}} \left |1010\right \rangle \xrightarrow{U_O^{(2)} \text{ 1 to 3}}\left |1000\right \rangle \xrightarrow{CNOT \text{ 2 to 3}} \left|1000 \right \rangle.\\
\end{align*}

As to $U_O^{(5)}$, first tensor $\left |1\right \rangle$ with the quantum state.
\begin{align*}
  \left |0000\right \rangle \xrightarrow{\otimes \left |1\right \rangle } \left |00001\right \rangle \xrightarrow{U_O^{(4)}} \left |00001\right \rangle \xrightarrow{CNOT \text{ 1 TO 2}} \left |00001\right \rangle \\
  \left |0001\right \rangle \xrightarrow{\otimes \left |1\right \rangle } \left |00011\right \rangle \xrightarrow{U_O^{(4)}} \left |00010\right \rangle \xrightarrow{CNOT \text{ 1 TO 2}} \left |00010\right \rangle \\
  \left |0011\right \rangle \xrightarrow{\otimes \left |1\right \rangle } \left |00111\right \rangle \xrightarrow{U_O^{(4)}} \left |00100\right \rangle \xrightarrow{CNOT \text{ 1 TO 2}} \left |00100\right \rangle \\
  \left |0111\right \rangle \xrightarrow{\otimes \left |1\right \rangle } \left |01111\right \rangle \xrightarrow{U_O^{(4)}} \left |01000\right \rangle \xrightarrow{CNOT \text{ 1 TO 2}} \left |01000\right \rangle \\
  \left |1111\right \rangle \xrightarrow{\otimes \left |1\right \rangle } \left |11111\right \rangle \xrightarrow{U_O^{(4)}} \left |11000\right \rangle \xrightarrow{CNOT \text{ 1 TO 2}} \left |10000\right \rangle.\\
\end{align*}

\section{The procedure of $U_B^{(7)}$  }
\label{a2}
Here, we give the transformation of the quantum state in $U_B^{(7)}$

$U_B^{(7)}$:
\begin{align*}
  \left |000000\right \rangle \xrightarrow{U_B^{(4)} \otimes U_B^{(4)}} \left |000000\right \rangle \xrightarrow{adder-plus-1} \left |000001\right \rangle \xrightarrow{adder} \left |001001\right \rangle \\
   \xrightarrow{Toffoli} \left |001001\right \rangle \xrightarrow{CNOT} \left |000001\right \rangle \xrightarrow{adder-minus-1} \left |000000\right \rangle\\
   \left |000001\right \rangle \xrightarrow{U_B^{(4)} \otimes U_B^{(4)}} \left |000001\right \rangle \xrightarrow{adder-plus-1} \left |000010\right \rangle \xrightarrow{adder} \left |010010\right \rangle \\
   \xrightarrow{Toffoli} \left |010010\right \rangle \xrightarrow{CNOT} \left |000010\right \rangle \xrightarrow{adder-minus-1} \left |000001\right \rangle\\
   \left |000011\right \rangle \xrightarrow{U_B^{(4)} \otimes U_B^{(4)}} \left |000010\right \rangle \xrightarrow{adder-plus-1} \left |000011\right \rangle \xrightarrow{adder} \left |011011\right \rangle \\
   \xrightarrow{Toffoli} \left |011011\right \rangle \xrightarrow{CNOT} \left |000011\right \rangle \xrightarrow{adder-minus-1} \left |000010\right \rangle\\
   \left |000111\right \rangle \xrightarrow{U_B^{(4)} \otimes U_B^{(4)}} \left |000011\right \rangle \xrightarrow{adder-plus-1} \left |000100\right \rangle \xrightarrow{adder} \left |100100\right \rangle \\
   \xrightarrow{Toffoli} \left |100100\right \rangle \xrightarrow{CNOT} \left |000100\right \rangle \xrightarrow{adder-minus-1} \left |000011\right \rangle\\
   \left |001111\right \rangle \xrightarrow{U_B^{(4)} \otimes U_B^{(4)}} \left |001011\right \rangle \xrightarrow{adder-plus-1} \left |001100\right \rangle \xrightarrow{adder} \left |101100\right \rangle \\
   \xrightarrow{Toffoli} \left |101101\right \rangle \xrightarrow{CNOT} \left |000101\right \rangle \xrightarrow{adder-minus-1} \left |000100\right \rangle\\
   \left |011111\right \rangle \xrightarrow{U_B^{(4)} \otimes U_B^{(4)}} \left |010011\right \rangle \xrightarrow{adder-plus-1} \left |010100\right \rangle \xrightarrow{adder} \left |110100\right \rangle \\
   \xrightarrow{Toffoli} \left |110110\right \rangle \xrightarrow{CNOT} \left |000110\right \rangle \xrightarrow{adder-minus-1} \left |000101\right \rangle\\
   \left |111111\right \rangle \xrightarrow{U_B^{(4)} \otimes U_B^{(4)}} \left |011011\right \rangle \xrightarrow{adder-plus-1} \left |011100\right \rangle \xrightarrow{adder} \left |111100\right \rangle \\
   \xrightarrow{Toffoli} \left |111111\right \rangle \xrightarrow{CNOT} \left |000111\right \rangle \xrightarrow{adder-minus-1} \left |000110\right \rangle\\
\end{align*}

\section{The adder-plus-$d$ gate.}
\label{a3}

We can use the Fourier gate to prepare the  adder-plus-$d$ gate as Fig.\ref{f16}
\begin{figure}[hbtp]
  \centering
  \includegraphics{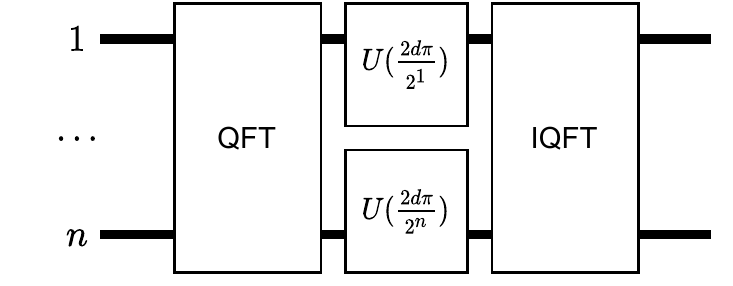}
  \caption{The Fourier adder-plus-$d$ gate.}
  \label{f16}
\end{figure}

An $n$-qubit adder-plus-$d$ gate ($n=\lceil \log \frac{N}{2} \rceil+1$ in our scheme) is composed of a quantum Fourier  transformation (QFT) gate, $n$ phase gates and an inverse quantum Fourier transformation (IQFT) gate. Given that the initial state is $\left | b_j \right \rangle$, we obtain
\begin{equation}
  \frac{1}{2^{n/2}} \mathop{\otimes} \limits_{k=0}^{n-1}(\left |0 \right\rangle+e^{2\pi ij \frac{2^k}{2^n}}\left |1 \right\rangle)
\end{equation}
after applying the QFT gate. Then, the action of phase gates $\mathop\otimes_{k=0}^{n-1} U(2d\pi\frac{2^k}{2^n})$,
 where $
  U(\lambda)=
  \begin{bmatrix}
    1 &0 \\
    0 &e^{i\lambda}
  \end{bmatrix}$,
is given by
\begin{equation}
  \frac{1}{2^{n/2}} \mathop{\otimes} \limits_{k=0}^{n-1}(\left |0 \right\rangle+e^{2\pi i(j+d) \frac{2^k}{2^n}}\left |1 \right\rangle),
\end{equation}
and finally, the state evolves into $\left | b_{j+d} \right \rangle$ after IQFT.

The Fourier adder-plus-$d$ gate has $O(n)$ depth and $O(n \log n)$ size. If we want to persue a smaller complexity, we can use the quantum carry lookahead adder (QLCA) method \cite{draper_logarithmic-depth_2006}. By adding $O(n)$ ancilla, QCLA can implement the adder with $O(\log n)$ depth and  $O(n)$ size.

\section{The recursion converter}
\label{a4}
In this section, we introduce the recursion converter presented by Plesch $et$ $al$. \cite{plesch_efficient_2010}. Although Plesch $et$ $al$. prove that the circuit depth has $O(N\log^2 N)$ depth and size, the circuit depth can decrease to $O(N\log N)$, because Saeedi $et$ $al$. has presented a scheme to decompose $n$-qubit Toffoli gates in linear depth.

In Plesch's scheme, the scheme is used to convert between the one-hot encoding state and the binary encoding state. Here by a slight change, the circuit can convert between the Edick state and the binary state. An $N=6$ qubit converter is shown as Fig. \ref{f16}.
\begin{figure}[hbtp]
  \centering
  \includegraphics{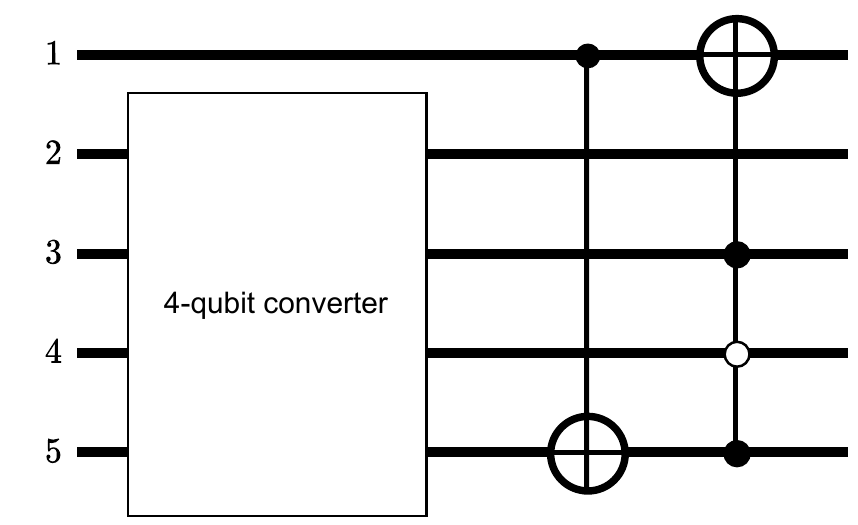}
  \caption{The converter with binary calculation.}
  \label{f16}
\end{figure}

Given that $N-2$ qubit converter $C_{N-1}$ has been prepared and we have already prepared $\left | \psi_E^{(N)} \right \rangle$. After applying $C_{N-1}$ on the last $N-2$ qubits, the quantum state becomes
\begin{equation}
  \alpha_{N-1} \left |1\right \rangle \left |0\right \rangle^{\otimes N-1-\lceil \log N \rceil} \left |b_{N-2}\right \rangle+ \left |0\right \rangle \sum_{i=0}^{N-2} \alpha_{i}   \left |0\right \rangle^{\otimes N-1-\lceil \log N \rceil} \left |b_i\right \rangle.
\end{equation}

After calculating the bitwise XOR $c=c_1 \cdots c_{\lceil \log N \rceil}=b_{N-1} \oplus b_{N-2} $, for all $i$ satisfying $c_i=1$, CNOT gates are implemented on the first qubit and the $N-\lceil \log N \rceil+i$ th qubit. Then, use the multiple qubit Toffoli gate to flip the first qubit if the last $\lceil \log N \rceil$ qubit is $\left |b_{N-1}\right \rangle$. Thus the final state becomes the binary state. In the original scheme \cite{plesch_efficient_2010}, we can replace $c$ with $ b_{N-1}$ to prepare $\left |\psi_B^{(N-1)}\right \rangle$ from $\left |\psi_O^{(N-1)}\right \rangle$
$\left |b_{N}\right \rangle$
\section{Generation of the binomial distribution state.}
\label{a5}

As Fig. \ref{f5} shown, we first implement $R_Y(\theta)^{\otimes N}$ on $\left|0\right \rangle ^{\otimes N}$, where $\theta=2\arcsin \sqrt{p}$ and $R_Y(\theta)=\begin{bmatrix} \cos \frac{\theta}{2} & -\sin \frac{\theta}{2}\\\sin \frac{\theta}{2} & \cos \frac{\theta}{2}\end{bmatrix}$ and we will obtain
\begin{equation}
  R_Y(\theta)^{\otimes N}(\left|0\right \rangle ^{\otimes N}) = \sum_{k=0}^N \sqrt{f(k,p)}\left |D_k^N  \right \rangle,
\end{equation}
 where $\left | D_k^{N} \right \rangle=\binom{N}{k}^{-\frac{1}{2}} \sum_{x \in \{0,1\}^N, wt(x)=k} \left |x \right \rangle$ is the Dicke state. Second, we implement the inverse gate of $U_{N,N-1}$ presented in \cite{bartschi_deterministic_2019} and the state becomes
\begin{equation}
  U_{N,N-1} \left |0 \right \rangle^{\otimes N-k} \left |1 \right \rangle ^{\otimes k} = \left | D_k^{N} \right \rangle
\end{equation}
for each $k \in \{0, 1, \cdots, N\}$, so $U_{N,N-1}^{-1}$ satisfies that
\begin{equation}
  U_{N,N-1}^{-1} \sum_{k=0}^N \sqrt{f(k,p)}\left |D_k^N  \right \rangle = \sum_{i=0}^{N} \sqrt{f(i, p)} \left |0 \right \rangle^{\otimes N-i} \left |1 \right \rangle ^{\otimes i}.
\end{equation}

$U_{N,N-1}$ can be decomposed to Split $\&$ Cyclic Shift gate ($SCS_{n,k}$) that is in the form of
\begin{equation}
  U_{N, N-1}:=\prod_{\ell=2}^{N-1}\left(S C S_{\ell, \ell-1} \otimes I^{\otimes N-\ell}\right) \cdot \left(I^{\otimes N-k-1} \otimes S C S_{N, N-1} \right),
\end{equation}
where $SCS_{n,k}$ satisfies that
\begin{align}
  \begin{aligned}
    \begin{array}{ll}
      S C S_{n, k}|0\rangle^{\otimes k+1} & =|0\rangle^{\otimes k+1}, \\
      S C S_{n, k}|0\rangle^{\otimes k+1-\ell}|1\rangle^{\otimes \ell} & =\sqrt{\frac{\ell}{n}}|0\rangle^{\otimes k+1-\ell}|1\rangle^{\otimes \ell}+\sqrt{\frac{n-\ell}{n}}|0\rangle^{\otimes k-\ell}|1\rangle^{\otimes \ell}|0\rangle, \\
      S C S_{n, k}|1\rangle^{\otimes k+1} & =|1\rangle^{\otimes k+1} .
      \end{array}
  \end{aligned}
\end{align}

\begin{figure}[htbp]
  \centering
  \begin{minipage}[t]{0.48\textwidth}
  \centering
  \includegraphics[width=6cm]{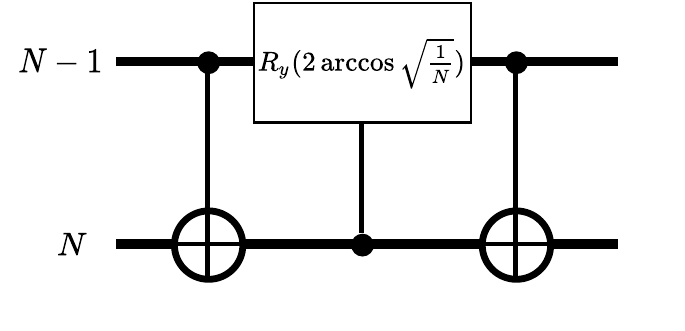}
  \caption{Two-qubit gates $SCS_2$.}
  \label{f6}
  \end{minipage}
  \begin{minipage}[t]{0.48\textwidth}
  \centering
  \includegraphics[width=6cm]{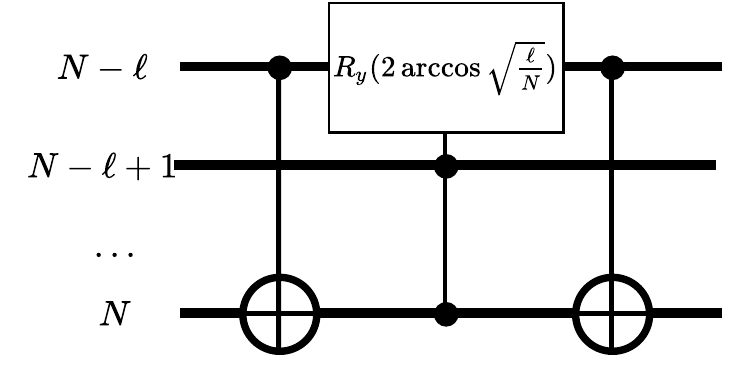}
  \caption{Three-qubit gates $SCS_3$.}
  \label{f7}
  \end{minipage}
\end{figure}

As proved in \cite{bartschi_deterministic_2019}, $SCS_{n,k}$ can be decomposed to 1 two-qubit gates and $k-1$ three-qubit gates shown as Fig. \ref{f6} and Fig. \ref{f7}.

Finally, we apply $U_B^{(N+1)}$ or $U_O^{(N+1)}$ to transform the Edick state to the one-hot encoding state or the binary encoding state \footnote{A extra ancilla qubits should added to $U_O^{(N+1)}$ if the one-hot encoding state is required.}, which is denoted as
\begin{equation}
  U_O^{(N+1)} (\sum_{i=0}^{N} \sqrt{f(i, p)} \left |0 \right \rangle^{\otimes N-i} \left |1 \right \rangle ^{\otimes i} \otimes \left | 1 \right \rangle)=\sum_{i=0}^{N} \sqrt{f(i, p)} \left |0 \right \rangle ^{\otimes N-i} \left | 1\right \rangle\left | 0\right \rangle^{\otimes i}
\end{equation}
\begin{equation}
  U_B^{(N+1)} (\sum_{i=0}^{N} \sqrt{f(i, p)} \left |0 \right \rangle^{\otimes N-i} \left |1 \right \rangle ^{\otimes i})=\sum_{i=0}^{N} \sqrt{f(i, p)}  \left | 0\right \rangle^{\otimes N-\log N} \left |b_i \right \rangle .
\end{equation}

\section{Proof in Sec. \ref{s1}}
\label{a6}
\subsection{The solution of Eq. \ref{e02}}
We use the binary $b_N=c_n c_{n-1}\cdots c_1$ to represent $N$. From Eq. \ref{e02}, we have
\begin{align*}
  d_o(c_n c_{n-1}\cdots c_1)&=d_o(c_n c_{n-1}\cdots c_{2} 0)\\&=d_o(c_n c_{n-1}\cdots c_{2})+2\\&=d_o(c_n c_{n-1}\cdots c_{3})+2*2\\&=d_o(c_n c_{n-1})+2(n-2)\\&=2n-1=2\lceil \log N \rceil-1
\end{align*}
\subsection{The solution of Eq. \ref{e03}}
We use the binary $b_N=c_n c_{n-1}\cdots c_1$ to represent $N$. From Eq. \ref{e03}, we have
\begin{align*}
  s_o(c_n c_{n-1}\cdots c_1)&=s_o(c_n c_{n-1}\cdots c_{2} 0)+c_1\\&=s_o(c_n c_{n-1}\cdots c_{2})+N\\&=s_o(c_n c_{n-1}\cdots c_{3})+N+\lfloor \frac{N}{2} \rfloor+c_1+c_2\\&=s_o(c_n c_{n-1})+\sum_{i=0}^{n-3} \lfloor \frac{N}{2^i} \rfloor +\sum_{i=1}^{n-2} c_i\\&==1+\sum_{i=0}^{n-3} \lfloor \frac{N}{2^i} \rfloor +\sum_{i=1}^{n-1} c_i<1+N+\log N
\end{align*}
\subsection{The solution of Eq. \ref{e04}}

Note $M=N-1$, $d(M)=d_b(N)$ and $s(M)=s_b(N)$, so $b_M=10^{\otimes (m+1)}$. We have
\begin{align*}
  d_b(N)&=d(10^{\otimes (m+1)})\\&=d(10^{\otimes (m)})+O(m)\\
  &=d(10^{\otimes (m-1)})+O(m)+O(m-1)\\
  &=d(10)+O(m^2)=O(\log^2 N)
\end{align*}

\begin{align*}
  s_b(N)&=2s(10^{\otimes (m)})+O(m)\\
  &=4s(10^{\otimes (m-1)})+O(m)+2O(m-1)\\
  &=2^{m}s(10)+\sum_i^{m-1} 2^i O(m-i)\\
  &=O(2^m)+km\sum_i^{m-1} 2^i-k\sum_{i=0}^{m-1} 2^i i\\
  &=O(2^m)+k2^{m+1}+O(m)=O(N)
\end{align*}

\subsection{The solution of Eq. \ref{e07}}
Note that $M=N-1$ , $d(M)=d_b(N)$. If we use the binary $b_M=c_n c_{n-1}\cdots c_1$ to represent $M$, then $O(\log N)=O(n)$. So we have
\begin{align*}
  d_b(N)&=d(c_n c_{n-1}\cdots c_1)\\&=d(c_n c_{n-1}\cdots c_{2} + c_1)+O(n)\\&=d(c_n c_{n-1}\cdots c_{3} + c_{2} \lor c_1)+O(n)+O(n-1)\\&=d(c_n c_{n-1}+c_{n-2}\lor \cdots \lor c_1)+\sum_{i=0}^{n-3} O(n-i) = O(\log^2 N)\\
\end{align*}

\subsection{The solution of Eq. \ref{e08}}
Note that $M=N-1$ , $s(M)=s_b(N)$. If we use the binary $b_M=c_n c_{n-1}\cdots c_1$ to represent $M$, then $O(\log N)=O(n)$. So we have
\begin{align*}
  s_b(N)&=s(c_n c_{n-1}\cdots c_1)\\&=2^1s(c_n c_{n-1}\cdots c_{2} + c_1)+O(n^2)\\&=2^2s(c_n c_{n-1}\cdots c_{3} + c_2 \lor c_1)+O(n^2)+2O((n-1)^2)\\&=2^{n-2}d(c_n c_{n-1}+c_{n-2}\lor \cdots \lor c_1)+\sum_{i=0}^{n-3} 2^i O((n-i)^2)\\&=O(2^n)+2^n\sum_{k=3}^{n}2^{-k}O(k^2)=O(2^n)-O(n^2)=O(N)\\
\end{align*}

\subsection{The solution of Eq. \ref{e05}}
Note that $M=N-1$ , $d(M)=d_b(N)$. If we use the binary $b_M=c_n c_{n-1}\cdots c_1$ to represent $M$, then $O(\log N)=O(n)$. So we have
\begin{align*}
  d_b(N)&=d(c_n c_{n-1}\cdots c_1)\\&=d(c_n c_{n-1}\cdots c_{2} 0)+c_1 O(n)\\&=d(c_n c_{n-1}\cdots c_{2})+(c_1+1) O(n)\\&=d(c_n c_{n-1}\cdots c_{3})+(c_1+1) O(n)+(c_2+1) O(n-1)\\&=d(c_n c_{n-1})+\sum_{i=1}^{n-2}(c_i+1) O(n-i+1) \\&<2 O(n^2)=O(\log^2 N)\\
\end{align*}

\subsection{The solution of Eq. \ref{e06}}
Note that $M=N-1$ , $s(M)=s_b(N)$. If we use the binary $b_M=c_n c_{n-1}\cdots c_1$ to represent $M$, then $O(\log N)=n$. So we have
\begin{align*}
  s_b(N)&=s(c_n c_{n-1}\cdots c_1)\\&=s(c_n c_{n-1}\cdots c_{2} 0)+c_1 O(n^2)\\&=2s(c_n c_{n-1}\cdots c_{2})+(c_1+1) O(n^2)\\&=2^2s(c_n c_{n-1}\cdots c_{3})+(c_1+1) O(n^2)+2(c_2+1)O((n-1)^2)\\&=2^{n-2}s(c_n c_{n-1})+\frac{1}{2}\sum_{i=1}^{n-2}2^i(c_i+1) O((n-i+1)^2) \\&<O(2^n)+\sum_{i=1}^{n-2}2^iO((n-i)^2)=O(2^n)+2^n\sum_{k=2}^{n-1}2^{-k}O(k^2)\\&=O(2^n)-O(n^2)=O(N)
\end{align*}
\end{appendices}

\end{document}